\documentclass [11pt,a4paper]{article}
\usepackage{jheppub}

\usepackage{graphicx}
\usepackage{color}
\usepackage{amsfonts,amsthm,amssymb,amsmath}
\usepackage{nicefrac}
\def\beq{\begin{equation}}
\def\eeq{\end{equation}}
\newcommand{\be}{\begin{eqnarray}}
\newcommand{\ee}{\end{eqnarray}}

\renewcommand{\texttt}{{}}

\def\bs{\begin{subequations}}
\def\es{\end{subequations}}

\def\Fc{\mathcal{F}}

\def\Hc{\mathcal{H}}

\def\Lc{\mathcal{L}}
\def\Mc{\mathcal{M}}

\def\Pc{\mathcal{P}}

\def\Rc{\mathcal{R}}

\def\con{\color{black}} %n for nera

\newcommand{\tia}[1]{}

\newcommand{\bea}{\begin{eqnarray}}
\newcommand{\eea}{\end{eqnarray}}
\newcommand{\beas}{\begin{eqnarray*}}
\newcommand{\eeas}{\end{eqnarray*}}
\newcommand{\bal}{\begin{aligned}}
\newcommand{\eal}{\end{aligned}}

\def\({\left(}
\def\){\right)}

\newcommand{\LF}{\left(}
\newcommand{\RF}{\right)}
\newcommand{\LT}{\left[}
\newcommand{\RT}{\right]}

\def\con{\color{black}} %n for nera

\newcommand{\pd}{\partial}

\newcommand{\cpd}{\nabla}

\newcommand{\const}{\mathrm{const}}

\title{Occurrence of exact \boldmath $R^2$ inflation 
in non-local UV-complete gravity}

\author[a,b]{Alexey S. Koshelev,}
\author[c,d]{Leonardo Modesto,}
\author[d]{Les\l{}aw Rachwa\l{} } 
\author[e,f]{and\\Alexei A. Starobinsky}
\affiliation[a]{Departamento  de F\'isica and Centro  de  Matem\'atica  e 
Aplica\c c\~oes (CMA-UBI),  Universidade  da  Beira  Interior,  6200  Covilh\~a, 
Portugal}
\affiliation[b]{Theoretische Natuurkunde, Vrije Universiteit Brussel, and The International Solvay Institutes, Pleinlaan 2, B-1050 Brussels, Belgium}
\affiliation[c]{Department of Physics, Southern University of Science and Technology, 
Shenzhen 518055, China, }
\affiliation[d]{Department of Physics \& Center for Field Theory and Particle Physics, \\
Fudan University, 200433 Shanghai, China}
\affiliation[e]{L. D. Landau Institute for Theoretical Physics RAS, Moscow 119334, Russian Federation}
\affiliation[f]{Kazan Federal University, Kazan 420008, Republic of Tatarstan, Russian Federation}
\emailAdd{alexey@ubi.pt}
\emailAdd{lmodesto@fudan.edu.cn}
\emailAdd{rachwal@fudan.edu.cn}
\emailAdd{alstar@landau.ac.ru}

\abstract{
The $R+R^2$, shortly named ``$R^2$'' (``Starobinsky'') inflationary model, represents a fully consistent example of a one-parameter inflationary scenario. This model has a ``graceful exit'' from inflation and provides a mechanism for subsequent creation and final thermalization of the standard matter. Moreover, it produces a very good fit of the observed spectrum of primordial perturbations. In the present paper we show explicitly that the $R^2$ inflationary spacetime is an exact solution of a range of weakly non-local (quasi-polynomial) gravitational theories, which provide an ultraviolet completion of the $R^2$ theory. These theories are ghost-free, super-renormalizable or finite at quantum level, and perturbatively unitary. Their spectrum consists of the graviton and the scalaron that is responsible for driving the inflation. Notably, any further extension of the spectrum leads to propagating ghost degrees of freedom. We are aimed at presenting a detailed construction of such theories in the so called Weyl basis. Further, we give a special account to the cosmological implications of this theory by considering perturbations during inflation. The highlight of the non-local model is the prediction of a modified, in comparison to a local $R^2$ model, value for the ratio of tensor and scalar power spectra $r$, depending on the parameters of the theory. The relevant parameters are under control to be successfully confronted with existing observational data. Furthermore, the modified $r$ can surely meet future observational constraints.
}


\begin{document}

\maketitle

%\tableofcontents

%\PACS 04.60.-m, 11.10.Lm, 11.25.Yb
% 04.60.-m	Quantum gravity
% 11.10.Lm	Nonlinear or non-local theories and models
% 11.25.Yb	M theory

%\date{\small April 7, 2016}

%%%%%%%%%%%%%%%%%%%%%%%%%%%%%%%%%%%%%%%%%%%%%%%%%%%%%%%%%%%%%%%%%%%%%%%%%%%%%%%%%%%%%%%%%
%%%%%%%%%%%%%%%%%%%%%%%%%%%%%%%%%%%%%%%%%%%%%%%%%%%%%%%%%%%%%%%%%%%%%%%%%%%%%%%%%%%%%%%%%
\clearpage
\section{Introduction}

The inflationary spacetime produced by a local $R+R^2$ gravity where $R$ is the Ricci scalar (``Starobinsky'' inflation, which we will 
refer to as ``$R^2$'' inflation\footnote{In the course of this paper terms ``$R^2$ inflationary model (spacetime, metric, etc.)'' or just ``$R^2$ 
model (inflation, gravity, etc.)'' refer to the local $R+R^2$ model and its solutions. Note also that this is a simplified form of the model 
studied in \cite{Starobinsky:1980te} in the limit $M\ll H$ using notations of that paper.}) \cite{Starobinsky:1980te,Starobinsky:1981vz,Starobinsky:1983zz} stays 
as a very successful model already for several decades. Theory-wise this model is the simplest extension to Einstein's general relativity (GR) with just one 
extra parameter (being the large dimensionless coefficient in front of the $R^2$ term in the action). The minimal number of parameters is a very elegant property of this model. Physics-wise this model features a ``graceful exit'' from inflation and provides a natural mechanism for the subsequent creation and final thermalization of the standard matter. Importantly, it produces a very good fit of the observed spectrum of primordial scalar perturbations 
and related parameters, such as the ratio of tensor to scalar power spectra $r$ \cite{Ade:2015xua,Ade:2015lrj,Array:2015xqh}.

Being good at the classical level, $R^2$ gravity by itself does not admit an ultraviolet-complete (UV-complete) quantum treatment since it is not renormalizable. Stelle has shown \cite{Stelle:1977ry,Stelle:1976gc} that adding the Weyl tensor squared term to the action makes it renormalizable, but then a tensor ghost appears. However, if the dimensionless coefficient in front of this term is of order of a unit and is much smaller than the one in front of the $R^2$ term, this ghost appears at very large values of the spacetime Riemann curvature close to the Planck scale and thus is much exceeding the curvature during the observable part of inflation which is 10 orders of magnitude less. Thus, we may use the $R^2$ model as an effective theory for description of inflation and post-inflationary evolution of the Universe. Anyway, if we believe that $R+R^2$ theory is the correct one in the inflationary era, and, at the same time, we assume renormalizability to be a guiding principle (so successful for all the other fundamental interactions), then we are forced to look for a completion of the local $R^2$ gravity in the UV regime \cite{Buchbinder:1992rb,Asorey:1996hz,Accioly:2002tz,Salles:2014rua}. Moreover, working in the quantum field theory framework we require the theory to be consistent at any perturbative loop order.  

In this paper we show explicitly that the $R^2$ inflationary metric is an exact solution to the equations of motion (EOM) for a large class of weakly non-local (quasi-polynomial) theories of gravity \cite{Tomboulis:1997gg,Tomboulis:2015esa,Tomboulis:2015gfa,Biswas:2005qr,Briscese:2013lna}. Actions of such theories are characterized by a polynomial structure in Riemannian curvatures with insertions of  fully covariant differential operators with infinite number of derivatives. An example operator can be written as $\Fc(\Box)$ with $\Box$ being the covariant d'Alembertian operator and $\Fc(z)$ an entire function. The crucial property assumed in constructions of this class of actions is that all such non-local operators are analytic in their arguments. That is these theories are non-local and hereafter we refer to them as  to ``non-local'' ones.\footnote{Analyticity of functions of differential operators implies that theories based on the pure inverse d'Alem\-ber\-tian \cite{Deser:2007jk} are strictly separated from our considerations.} 

The consideration of non-local gravity theories in its current form was initiated by Tomboulis \cite{Tomboulis:1997gg} in an attempt to find a UV-complete gravitational theory. This line of research propagated to the construction of a quasi-polynomial theory \cite{Briscese:2013lna}, that realizes a unitary completion of GR. There exists a subclass of theories, such that at quantum level they are super-renormalizable or finite and moreover ghost-like perturbative degrees of freedom do not show up \cite{Accioly:2002tz,Modesto:2011kw,Modesto:2014lga, Talaganis:2014ida,Dona:2015tra}. At  the level of classical solutions of these theories the Newtonian potential turns out to be a regular at short distances (high energies) with a universal constant limit at zero distance \cite{Stelle:1977ry,Modesto:2010uh,Biswas:2011ar,Biswas:2013cha,Bambi:2013gva,Calcagni:2013vra,Modesto:2014eta,Frolov:2015bta,Frolov:2015bia,Frolov:2015usa,Edholm:2016hbt}. The spectrum of these theories consists solely of the graviton and the scalaron -- the quantum scalar particle corresponding to the gravitational scalar degree of freedom arising in $f(R)$ garvity with $\frac{d^2f}{dR^2}\not\equiv 0$. It was shown in \cite{Briscese:2013lna} that any extension of the spectrum beyond this leads to a propagation of ghosts. 
Moreover as opposite to GR or the local quadratic gravity in which understanding of stability has taken a long-time effort \cite{Arnowitt:1960es,Abbott:1981ff,Starobinsky:1987zz,Muller:1987hp,Deser:2007vs}, the Minkowski vacuum is naturally stable in non-local gravity \cite{Biswas:2011ar,Modesto:2015lna, Modesto:2015foa}.

Account of non-local gravity effects, namely, backreaction of one-loop corrections of quantized matter fields (both massless and massive) on the evolution of a FLRW background and metric perturbations, had been already a crucial part of the model in \cite{Starobinsky:1980te} since it produced 
decay of scalarons into pairs of particles and antiparticles of all quantum matter fields after the end of inflation, thus, providing the final 
transition of the Universe to the radiation dominated stage.\footnote{In \cite{Starobinsky:1980te}, the final results of corresponding calculations were presented 
only. The detailed calculations of particle creation using the formalism of $\alpha,\beta$-coefficients of the Bogolyubov transformation can 
be found in \cite{Starobinsky:1981vz}. Note also that decay of scalarons into pairs of gravitons appears to be strongly suppressed \cite{Starobinsky:1981zc} that it is very 
important for the viability of the model, too.} After that cosmological considerations of non-local gravity were re-initiated in \cite{Aref'eva:2004vw}, where non-local scalar field Lagrangian coming from string field theory (SFT) \cite{Witten:1986qs,Vladimirov:1994wi,Arefeva:2001ps} was minimally coupled to GR in an attempt to describe Dark Energy. Notice that SFT natively produces the non-local structures essentially similar to those suggested ad hoc in \cite{Tomboulis:1997gg}. Moreover, SFT by its construction is unitary and UV-complete. The first explicit gravitational action that includes non-local terms involving the scalar curvature and an exact analytic solution for a non-singular bounce in this model were described in \cite{Biswas:2005qr}. Further cosmologically oriented explorations of this gravitational action can be found in e.g. \cite{Biswas:2013dry,Chialva:2014rla}.\footnote{We also point out related studies addressing aspects of non-local gravity theories while having inverse powers of the d'Alembertian as well \cite{Barvinsky:2003kg,Barvinsky:2011hd,Barvinsky:2014lja,Conroy:2014eja}.} It was shown in \cite{Koshelev:2013lfm} that the most general quadratic in curvature non-local gravity action may contain $R^2$, $R_{\mu\nu}^2$ and $R_{\mu\nu\alpha\beta}^2$ terms sandwiched by non-local differential operators. As long as we are focusing on inflation, which is a nearly de Sitter (dS) expansion phase, curvature squared terms give the essence of the underlying physics. A rigorous proof of this fact and a perturbative treatment of such theories around maximally symmetric spacetimes are accumulated in \cite{Biswas:2016etb}.

In \cite{Craps:2014wga} for the first time the $R^2$ inflation was considered in a framework of non-local gravity of the type $R+R\Fc(\Box)R$. It was shown that predictions for power spectra and their ratio for tensor and scalar perturbations remain as in a local $R+R^2$ model. This can be understood as follows. The scalar curvature squared piece does not generate tensor modes thus leaving them as they were in GR. Focusing on the nearly dS phase effectively eliminates the non-locality as all background quantities are covariantly constant.

In the present paper we perform an analysis similar to the one in \cite{Craps:2014wga} in a presence of the non-local Weyl tensor squared term in the action. We thoroughly study classical perturbations and their quantization. As the net result we deduce power spectra for tensors and scalars and compute their ratio. It is expected that this time there will be extra tensor modes contributions generated from new terms with tensorial structure in the action and that the ratio $r$ will be modified. This is the novel result of the paper, which gives the theory the predictive power as now it can be checked with existing or future observational data about cosmological perturbations.

The paper is organized as follows. In the second section we introduce a generic class of weakly non-local theories, we evaluate the propagator, and we shortly remind the power counting analysis of
super-renormalizability in these theories. 
In particular, we construct a theory in the ``Weyl basis'', which is slightly different from the one proposed in \cite{Briscese:2013lna}. The Weyl basis proves useful for considering Friedmann-Lema\^itre-Robertson-Walker (FLRW) solutions.
In the third section we explicitly show that the $R^2$ inflationary metric is 
an exact solution of the constructed weakly non-local theory. 
In the fourth section we compute the second order variation of the action around the nearly dS phase of inflation and derive cosmological parameters. In the fifth section we conclude by summarizing the results and promoting open questions. Appendices contain notations used throughout the paper as well as additional derivations and verifications of some computations.

%%%%%%%%%%%%%%%%%%%%%%%%%%%%%%%%%%%%%%%%%%%%%%%%%%%%%%%%%%%%%%%%%%%%%%%%%%%%%%%%%%%%%%%%%%%

\section{General multidimensional non-local gravity}
The Lagrangian density of the most general $D$-dimensional theory weakly non-local (quasi-polynomial) and quadratic in curvature reads 
\cite{Krasnikov:1987yj,Tomboulis:1997gg,Khoury:2006fg,Biswas:2011ar,Modesto:2011kw, Modesto:2014lga,Alexander:2012aw,Briscese:2012ys,Briscese:2013lna,Modesto:2013jea,Calcagni:2014vxa},
\begin{eqnarray}
&& 
\mathcal{L}_{\rm g} =
 \frac12 \kappa_{D}^{-2} \, \sqrt{|g|} 
\left[ {\bf R} 
+
{\bf R} \, 
 \gamma_0(\Box)
 {\bf R} 
 + {\bf Ric} \, 
\gamma_2(\Box)
 {\bf Ric} 
+ {\bf Riem}  \, 
\gamma_4(\Box)
{\bf Riem} 
+ V \, 
\right] \, .
\label{pregravityG}
\end{eqnarray}
We refer the reader to appendix~\ref{ap:notation} for notations and conventions used in this paper.
Here we notice that $\Box = g^{\mu\nu} \nabla_{\mu} \nabla_{\nu}$ is the covariant d'Alembertian operator while operators $\gamma_i(\Box)$ are termed form-factors. These form-factors are analytic functions of their argument. Physically speaking the proper argument should be written as 
$
z\equiv\Box_\Mc\equiv\Box/\Mc^2\, ,$
where $\Mc$ is an invariant fundamental mass scale in our theory.

Since the main problem of the quantum Einstein gravity (which remains in the local $R+R^2$ quantum gravity, too) is the appearance of
the massive tensor ghost with mass $\sim {\kappa_D}^{-1}$ at the one-loop level \cite{Stelle:1976gc},  any UV-complete  non-local generalization of gravity 
has to begin below this scale. Thus, we assume that ${\cal M}\lesssim M_P\equiv\kappa_D^{-1}$. On the other hand, observational data 
on the amplitude and slope of the primordial power spectrum of scalar (density) perturbations \cite{Ade:2015xua,Ade:2015lrj,Array:2015xqh}, if we want to describe them using the 
$R^2$ inflationary model, require the dimensionless coefficient in front of the $R^2$ term in the action to be very large and equal to 
$\approx 5\times 10^8 $ \cite{Starobinsky:1983zz}. As a result, the rest mass $M$ of the scalaron should be $$M=1.3 \times 10^{-5} \, \frac{55}{N_0} M_P=3.2 \times 10^{13}\, \frac{55}{N_0} \, {\rm GeV}\, ,$$
 where $N_0$ is the number of e-folds from the end of inflation corresponding to the pivot point 
$k=k_0=0.05$ Mpc$^{-1}$ in the measurements of the scalar power spectrum (here $k$ is the spatial momentum). Moreover, the same data show that the terms higher order 
in $R$, namely $R^n$ with $n>2$, should be strongly suppressed during the observable part of inflation which occurs in the range 
$R= (4 - 240)M^2$  (see Sec.~\ref{sec32} below) \cite{Huang:2013hsb}, so that gravity has approximately scale-free behaviour in some range of 
Ricci curvatures exceeding $M^2$. Thus, the cosmological data exclude the ``naive'' conjecture ${\cal M}\sim M$, and we have to assume 
the following hierarchy of scales:
\begin{equation}
M\ll {\cal M}\lesssim M_P \, .
\end{equation}
This phenomenological hierarchy suggests that the microscopical origins of inflation and non-local UV-completion of gravity are not the same.
It is important to emphasize that this hierarchy between Ricci scalar
and Weyl type corrections to general relativity is not destroyed by
quantum gravitational corrections, at least in the first order, as our
calculation of the generated graviton spectrum (finite and small) presented below shows.
We set $\Mc$ to unit hereafter unless stated otherwise.

We can rewrite the theory making use of a more compact notation introducing a tensorial form-factor, namely
\begin{eqnarray}
\mathcal{L}_{\rm g} &=& %-  
\frac12 \kappa_{D}^{-2} \, \sqrt{|g|} 
\left( {\bf R} 
+ {\bf Riem}  \, 
{\bf \gamma}(\Box)
{\bf Riem} 
+ {V} 
\right)\equiv  \nonumber \\
&\equiv& 
  \frac12 \kappa_{D}^{-2} \, \sqrt{|g|} 
\left(    R 
+ {R}_{\mu\nu\rho\sigma}  
\gamma(\Box)^{\mu \nu \rho \sigma}_{\alpha \beta \gamma \delta} \, 
{R}^{\alpha\beta\gamma\delta} 
+ { V} 
\right)\,  , 
\label{gravityG} %\\
\end{eqnarray}
where the operator in between the Riemann tensors is 
\begin{eqnarray*}
\gamma(\Box)^{\mu \nu \rho \sigma , \alpha \beta \gamma \delta} 
\equiv g^{\mu\rho} g^{\alpha\gamma} g^{\nu \sigma} g^{\beta \delta} \gamma_0(\Box)
+ g^{\mu\rho} g^{\alpha\gamma} g^{\nu\beta} g^{\sigma\delta} \gamma_2(\Box) 
+ g^{\mu\alpha} g^{\nu\beta} g^{\rho \gamma} g^{\sigma \delta} \gamma_4(\Box)  \, . 
\end{eqnarray*}
The theory consists of an Einstein-Hilbert (EH) term, a kinetic weakly non-local operator quadratic in the curvature, and a local potential-like term ${V}$ made of the following three sets of operators, 
\begin{eqnarray}
 {V} &=& 
 \sum_{j=3}^{{\rm N}+2} \sum_{k=3}^{j} \sum_i c_{k,i}^{(j)} \left( \nabla^{2(j-k)} {\cal R}^k \right)_i
 +
 \sum_{j={\rm N}+3}^{\gamma+{\rm N}+1} \sum_{k=3}^{j} \sum_i d_{k,i}^{(j)} \left(\nabla^{2(j-k)} {\cal R}^k \right)_i
 \nonumber\\
 &+&
 \sum_{k=3}^{\gamma +{\rm N}+2}
 \sum_i s_{k,i} \, \left(  \nabla^{2 (\gamma + {\rm N}+2 -k )} \, {\cal R}^k \right)_i \, ,  
 \nonumber 
 \end{eqnarray} 
where the operators in the third set are called ``killers'', because they are crucial in making the theory finite in any 
 dimension.
 The coefficients
 $c_{k,i}^{(j)}$, $d_{k,i}^{(j)}$, $s_{k,i}$ are running or non RG-running coupling constants, while the tensorial structures of terms have been neglected. With symbol ${\cal R}$ we generally denote one of the above curvature tensors.
 The integer parameter
$\gamma$ will be defined shortly. The capital $\rm{N}$ is defined to be the following function of the spacetime dimension $D$: $2 \mathrm{N} + 4 = D$.
The form-factors $\gamma_i(\Box)$ are defined in terms of exponentials of entire functions $H_\ell(\Box)$ ($\ell=0,2$), namely 
\begin{eqnarray}
&& 
\gamma_0(\Box) = - \frac{(D-2) ( e^{H_0(\Box)} -1 ) + D ( e^{H_2(\Box)} -1 )}{4 (D-1) \Box} + \gamma_4(\Box) \,  , 
\label{gamma2}   \qquad %
\\&&
 \gamma_2(\Box) = \frac{e^{H_2(\Box)} -1 }{\Box} - 4 \gamma_4(\Box) \, ,
\label{gamma0}
\end{eqnarray}
while $\gamma_4(\Box)$ stays arbitrary. It is only constrained by renormalizability of the theory to have the same asymptotic polynomial UV behaviour as the other two form-factors $\gamma_\ell(\Box)$ ($\ell=0,2$). The minimal choice 
compatible with unitarity and super-renormalizability corresponds to  $\gamma_4(\Box) =0$.

As a matter of fact we can also add other operators quadratic in the curvature and equivalent to the above operators
up to interaction vertices. These operators correspond to a different ordering in introducing the form-factors 
in between of the Riemann, Ricci, and scalar curvatures. We name these operators ``terminators''
to distinguish them from the ``killer operators'' present in the potential-like term ${ V}$. 
Such non-local operators can be crucial in making the theory finite \cite{Kuzmin:1989sp}, if 
we do not introduce any local (or non-local) potential-like terms $V$ higher than quadratic in the curvature.
The non-local terminators, if expressed through entire functions do not affect the unitarity. Some examples of terminators are: 
$$
R \nabla_\alpha \frac{e^{H_{3}(\Box)} -1}{\Box^2}\nabla^\alpha R\, ,~
R_{\mu\nu}\nabla_\alpha\nabla_\beta \frac{e^{H_{4}(\Box)} -1}{\Box^3} \nabla^{\alpha} \nabla^{\beta}R^{\mu\nu}\, ,~  
  R_{\mu\nu\rho\alpha} \frac{e^{H_{5}(\Box)} -1}{\Box^2} \nabla_{\beta}\nabla^{\alpha} 
 R^{\mu\nu\rho\beta} 
\, , \, \dots
$$

Finally, the entire functions $V^{-1}_{\ell}(z) \equiv \exp (H_{\ell}(z))$  ($\ell=0,2$) in the action 
satisfy the following general conditions \cite{Tomboulis:1997gg}:
\begin{enumerate}
\renewcommand{\theenumi}{(\roman{enumi})}
\item 
$V^{-1}_{\ell}(z)$ is real and positive on the real axis and it has no zeros on the 
whole complex plane $|z| < + \infty$. This requirement implies that there are no 
gauge-invariant poles other than the physical pole of transverse massless graviton;
\item
$|V^{-1}_{\ell}(z)|$ has the same asymptotic behaviour along the real axis at $\pm \infty$; 
\item 
there exists $0<\varphi<\pi/2$, such that asymptotically
\begin{eqnarray*}
&& 
|V^{-1}_{\ell}(z)| \rightarrow | z |^{\gamma + \mathrm{N}+1}\, ,\: \text{ when } \:|z|\rightarrow + \infty\: \text{ with the integer parameter} \,\,\gamma\,\, \text{satisfying: }
\\
&& \text{(a) }\:  \gamma\geqslant \frac{D_{\rm even}}{2} \, , 
\text{ (b) }\: \gamma\geqslant \frac{D_{\rm odd}-1}{2} \, ,
\end{eqnarray*}
respectively in even and odd dimension for the complex values of $z$ in the conical regions defined by: 
$$ - \varphi < {\rm arg} z < + \varphi \, ,  
\quad  \pi - \varphi < {\rm arg} z < \pi + \varphi\, .$$
\end{enumerate}
The last condition is necessary to achieve the maximum convergence of the theory in
the UV regime.  
The necessary asymptotic behaviour is imposed not only on the real axis, but also on the conical regions, that surround it.  
In Euclidean spacetime, the condition (ii) is not strictly necessary, if (iii) applies.

An example of an exponential form-factor $\exp H_\ell(z)$ compatible with (i)-(iii), 
and the guiding principles of quantum field theory (locality of counterterms) is \cite{Tomboulis:1997gg}:
\begin{eqnarray}
 e^{H_\ell(z)}
&=&  e^{\frac{a}{2} \left[ \Gamma \left(0, p(z)^2 \right)+\gamma_E  + \log \left( p(z)^2 \right) \right] } =\nonumber\\
 &=& 
e^{a \frac{\gamma_E}{2}} \,
\sqrt{ p(z)^{2 a}}
\left\{ 
1+ \left[ \frac{a \, e^{-p(z)^2}}{2 \,  p(z)^2} \left(  1 
+ O \left(   \frac{1}{p(z)^2}  \right)    \right) + O \left(e^{-2 p(z)^2} \right)  \right] \right\} \, , 
 \label{Tomboulis}
\end{eqnarray}
where the last equality 
is correct only on the real axis. In the formula above $a$ is a positive integer,
 $\gamma_E \approx 0.577216$ is the Euler-Mascheroni constant and 
$
\Gamma(0,z) = \int_z^{+ \infty}  d t \, e^{-t} /t 
$ 
is the incomplete gamma function with its first argument vanishing (notice that it follows from the theory of special functions that
$\Gamma(0,z)=\mathrm{Ei}(1,z)$). 
The  polynomial $p(z)$ of degree $\gamma +\mathrm{N}+1$ is such that $p(0)=0$, which gives the correct low energy limit of our theory (coinciding with GR).
This entire function has asymptotic polynomial behaviour $z^{a(\gamma + {\rm N} +1)}$ 
in a conical region around the real axis with angular opening 
$\varphi= \pi/(4 (\gamma + \mathrm{N} + 1))$. For $\gamma =0$ and $\rm{N}=0$ we have the maximal conical region characterized by the opening angle $\varphi = \pi/4$.

\subsection{Propagator and unitarity} \label{propagatorMink}
Splitting the spacetime metric into the flat Minkowski background and the fluctuation $h_{\mu \nu}$ 
defined by $g_{\mu \nu} =  \eta_{\mu \nu} + \kappa_D \, h_{\mu \nu}$,
we can expand the action (\ref{pregravityG}) to the second order in $h_{\mu \nu}$.
The result of this expansion together with the usual harmonic gauge fixing term reads \cite{Accioly:2002tz}
$$
\mathcal{L}_{\rm quad} + \mathcal{L}_{\rm GF} = \frac12 h^{\mu \nu} \mathcal{O}_{\mu \nu, \rho \sigma}  h^{\rho \sigma}\, ,
$$
where the operator $\mathcal{O}$ is made out of two terms, one coming from the quadratic part of (\ref{pregravityG})
and the other from the following gauge-fixing term,
$$
\mathcal{L}_{\rm GF}  = \xi^{-1}  \partial^{\nu}h_{\mu \nu} w(\Box) \partial_{\rho}h^{\rho \mu}\, ,
$$
and $w( \Box)$ is a weight functional \cite{Stelle:1976gc,Buchbinder:1992rb}.
The d'Alembertian operator in $\mathcal{L}_{\rm quad}$ and the gauge fixing term must be conceived on %relative 
the flat spacetime. 
Inverting the operator $\mathcal{O}$ \cite{Accioly:2002tz} and making use of the
form-factors (\ref{gamma2}) and (\ref{gamma0}), we find the 
two-point function in the harmonic gauge ($\partial^{\mu} h_{\mu \nu} = 0$) and in momentum space,
\begin{eqnarray}
\mathcal{O}^{-1} = - \frac{\xi (2P^{(1)} + \bar{P}^{(0)} ) }{2 p^2 \, w(- p^2)} 
- \left( 
\frac{P^{(2)}}{p^2   e^{H_2(-p^2)} }
\label{propagator} 
- \frac{P^{(0)}}{  \left( D-2 \right)p^2   e^{H_0(-p^2)}} \right) \,  .
\end{eqnarray}
Here $p_\alpha$ is the 4-momentum and $p^2=p_\alpha p^\alpha$.
Above we omitted the tensorial 
indices for the propagator $\mathcal{O}^{-1}$ and the projectors $\{ P^{(0)}$, $P^{(2)}$, $P^{(1)}$, $\bar{P}^{(0)}\}$ 
defined by
\cite{Accioly:2002tz,VanNieuwenhuizen:1973fi}
\begin{eqnarray*}
 P^{(2)}_{\mu \nu, \rho \sigma}(p) &=&   \frac{1}{2}( \theta_{\mu \rho} \theta_{\nu \sigma} +
 \theta_{\mu \sigma} \theta_{\nu \rho} ) -  \frac{1}{D-1} \theta_{\mu \nu} \theta_{\rho \sigma}\, ,
 \nonumber \\
P^{(1)}_{\mu \nu, \rho \sigma}(p) &=&   \frac{1}{2} \left( \theta_{\mu \rho} \omega_{\nu \sigma} +
 \theta_{\mu \sigma} \omega_{\nu \rho}  +
 \theta_{\nu \rho} \omega_{\mu \sigma}  +
  \theta_{\nu \sigma} \omega_{\mu \rho}  \right)  \,  , \nonumber   \\
 P^{(0)} _{\mu\nu, \rho\sigma} (p) &=&    \frac{1}{D-1}\theta_{\mu \nu} \theta_{\rho \sigma}\, ,~ 
\bar{P}^{(0)} _{\mu\nu, \rho\sigma} (p) =  \omega_{\mu \nu} \omega_{\rho \sigma}\, ,\nonumber\\
\theta_{\mu \nu} &=&  \eta_{\mu \nu} - \frac{p_{\mu } p_{\nu }}{p^2}\, ,~
\omega_{\mu \nu } = \frac{p_{\mu} p_{\nu}}{p^2}\, .
\end{eqnarray*}
We also have replaced $\Box \rightarrow -p^2$ in the quadratic action, so going to the momentum space.

The propagator (\ref{propagator}) is the most general one compatible with unitarity describing a spectrum without any other degree of freedom besides the graviton field. In relation to unitarity issue, the optical theorem here is trivially satisfied, namely 
 \begin{equation} 
 2 \, {\rm Im} \left\{  T(p)^{\mu\nu} \mathcal{O}^{-1}_{\mu\nu, \rho \sigma} T(p)^{\rho \sigma} \right\} = 2  \pi \, {\rm Res} \left\{  T(p)^{\mu\nu} \mathcal{O}^{-1}_{\mu\nu, \rho \sigma} T(p)^{\rho \sigma} \right\} \big|_{p^2 = 0}> 0\, ,
 \end{equation}
 where $T(p)^{\mu\nu}$ is the most general conserved energy tensor written in Fourier space \cite{Accioly:2002tz}. 

The most general theory compatible with optical theorem at tree-level will contain also a scalar particle, the scalaron. Further extensions of the physical spectrum will inevitably introduce real ghost \cite{Briscese:2013lna}. We obtain this most general theory provided we make the following replacement,
\begin{eqnarray}
e^{H_0(\Box)} \, \rightarrow \, e^{H_0(\Box)} \left(  1 - \frac{\Box}{M^2}   \right)\, ,
\label{gammaSwfscalaron}
\end{eqnarray} on the level of definition of the  form-factor $\gamma_0(\Box)$ in \eqref{gamma2}.  In this case the gauge-invariant part of the propagator for all modes contained in the fluctuation field $h_{\mu\nu}$ is slightly different, namely

\begin{eqnarray}
\mathcal{O}^{-1} =
- \left( 
\frac{P^{(2)}}{p^2   e^{H_2(-p^2)} }
\label{propagators} 
- \frac{P^{(0)}}{  \left( D-2 \right)p^2   e^{H_0(-p^2)}\left(1+\frac{p^2}{M^2}\right)} \right) \,  .
\end{eqnarray}
In this kind of theories we really need the asymptotically polynomial behaviour of the form-factors. Moreover to have renormalizability the degree of the asymptotic polynomial appearing in the  entire function $e^{H_0(\Box)}$ with the scalar projector $P^{(0)}$ must be smaller by one than the corresponding one in $e^{H_2(\Box)}$, that is it must be given by $\gamma+{\rm N}$. We see that $M$ is the mass of the scalaron. The version of the optical theorem satisfied here is slightly modified \cite{Briscese:2013lna} and reads
 \begin{eqnarray} 
 2 \, {\rm Im} \left\{  T(p)^{\mu\nu} \mathcal{O}^{-1}_{\mu\nu, \rho \sigma} T(p)^{\rho \sigma} \right\} = 2  \pi \, {\rm Res} \left\{  T(p)^{\mu\nu} \mathcal{O}^{-1}_{\mu\nu, \rho \sigma} T(p)^{\rho \sigma} \right\} \big|_{p^2 = -M^2}> 0\,.
 \end{eqnarray}
 This is the theory that we will analyze deeply in the course of this paper.

\subsection{Simplified power counting}
\label{gravitonpropagator}
We now review the power counting analysis of the quantum divergences. 
In the UV regime, 
the above propagator (\ref{propagator}) in momentum space 
schematically scales as 
\begin{eqnarray}
\mathcal{O}^{-1}(p) \sim \frac{1}{p^{2 \gamma +D} } \, . 
\label{OV} 
\end{eqnarray}
The vertices can be collected in different sets, that may or not involve 
the entire functions $\exp (H_\ell(z))$. 
However, to find a bound on the quantum divergences it is sufficient to concentrate on
the leading operators in the UV regime. 
These operators scale as the propagator giving the following 
upper bound on the superficial degree of divergence $d$ of any graph $G$ \cite{Modesto:2011kw}, 
\begin{eqnarray}
d(G)=DL+(V-I)(2 \gamma + D)
\end{eqnarray} 
in a spacetime of even or odd dimensionality $D$. We simplify this to
\begin{eqnarray} 
&&
d(G) =  D - 2 \gamma  (L - 1)    \, .
\label{even}
\end{eqnarray}
In 
(\ref{even}),  we used the topological relation between the number of vertices $V$, internal lines $I$ and the
number of loops $L$: $I = V + L -1$. 
Thus, if $\gamma > D/2$, only 1-loop divergences survive in the theory.  
Therefore, 
the theory is super-renormalizable \cite{Krasnikov:1987yj,Alebastrov:1973vw,Alebastrov:1973np} 
and only a finite number of operators of mass dimension up to $[{\rm mass}]^D$ has to be
included in the action in an even dimension to secure renormalizability. The conclusions obtained here are valid not only  for theories with a graviton in the spectrum,  but also for theories, where we have an additional scalar particle.

\subsection{Theory in Weyl basis}\label{FiniteS}
In this section we consider a different gravitational action written in the Weyl basis, which is equivalent to the previous one \eqref{pregravityG} for 
everything, which concerns  unitarity (the propagator is given again by (\ref{propagator})) and super-renormalizability
or finiteness.
The Lagrangian density reads,
\begin{equation}
\mathcal{L}_{\rm C} =   \frac12 \kappa_D^{-2} \sqrt{|g|}\Big[ {\bf R} +
 {\bf C} 
  \gamma_{\rm C} (\Box) {\bf C}
 + {\bf R}  \gamma_{\rm S}(\Box) {\bf R }
 + {\bf Riem} \,  \gamma_{\rm R}(\Box) {\bf Riem} + { V_g}({\bf C})
 \Big]  \,  .
\label{TWeyl}
\end{equation}
Here the form-factors are found to be,
\begin{equation*}
\gamma_{\rm C}(\Box) =  - \frac{D-2}{4}  \gamma_2(\Box) \, , ~ 
\gamma_{\rm S}(\Box) = \gamma_0(\Box)  +  \frac{1}{2(D-1)}  \gamma_2(\Box) \, , ~\gamma_{\rm R}(\Box) 
= \gamma_4(\Box) + \frac{D-2}{4}  \gamma_2(\Box) \, ,
\end{equation*}
where all the form-factors $\gamma_{\ell}(\Box)$ ($\ell=0,2,4$) are defined in (\ref{gamma2}) and (\ref{gamma0}). Expressing $\gamma_{\rm C}(\Box)$, $\gamma_{\rm S}(\Box)$ and $\gamma_{\rm R}(\Box)$ via entire functions $H_0(\Box)$, $H_2(\Box)$ and $\gamma_4(\Box)$
 we find 
\begin{eqnarray}
\gamma_{\rm C} &=&  \frac{(D-2) \left(-e^{H_2(\Box)}+4 \gamma _4(\Box) \Box+1\right)}{4 \Box}\, ,\\
 \gamma_{\rm S} &=& \frac{(2-D) \left(e^{H_0(\Box)}+e^{H_2(\Box)}-2\right)+4 \gamma _4(\Box) (D-3) \Box }{4 (D-1) \Box}
\, , \\
\gamma_{\rm R} &=& 
\frac{(D-2) \left(e^{H_2(\Box)}-1\right)-4 \gamma _4(\Box) (D-3) \Box}{4 \Box}\, .
\end{eqnarray}
{For the sake of simplicity we can assume 
$\gamma_{\rm R}(\Box) =0$ , then the theory (\ref{TWeyl}) reduces to}
\begin{eqnarray}
&& { \mathcal{L}_{\rm C} =   \frac12 \kappa_D^{-2} \sqrt{|g|}\Big[ {\bf R} +
 {\bf C} 
  \gamma_{\rm C} (\Box) {\bf C}
  + {\bf R}  \gamma_{\rm S}(\Box) {\bf R }
  + {V_g}({\bf C})
 \Big] }  \,  ,
 \label{TWeyl2} 
 \\
&&
\gamma_{\rm C} =   \frac{D-2}{4(D-3)}  \frac{e^{H_2(\Box)} -1}{\Box} \, , \quad 
\gamma_{\rm S} = - \frac{D-2}{4(D-1)}  \frac{e^{H_0(\Box)} -1}{\Box} \, .
\nonumber 
\end{eqnarray}
If a potential-like term ${ V_g}({\bf C})$, that we can always built up with only Weyl tensor, is included in the action,  this theory likely turns out to be UV-finite at the quantum level, because all remaining divergences at one loop level can be consistently cancelled.

In $D=4$ it is enough to include in $V$ a term
made out of two Weyl killer operators to end up with a finite quantum gravitational theory at any perturbative order in the loop expansion. For example we can choose the following two operators,
\begin{equation}
{ V_g}({\bf C}) = s^{(1)}_{\rm C}  C_{\mu\nu\rho\sigma} C^{\mu\nu\rho\sigma} \Box ^{\gamma -2} C_{\alpha \beta \gamma \delta}
C^{\alpha \beta \gamma \delta} +
s^{(2)}_{\rm C}  C_{\mu\nu\rho\sigma} C^{\alpha \beta \gamma \delta} \Box ^{\gamma -2} C_{\alpha \beta \gamma \delta}
C^{\mu\nu\rho\sigma} \, . 
\label{WeylKiller}
\end{equation} 
Since the beta-functions can only be linear in the front coefficients $s^{(1)}_{\rm C}$ and $s^{(2)}_{\rm C}$, 
we can always find a solution for $\beta_{\bf R^2} =0$ and 
$\beta_{\bf Ric^2} =0$ at any energy scale. (Actually the beta-functions here do not depend on any scale). We proved the multi-loop Feynman diagrams to be finite and, therefore, the beta functions are actually one-loop exact (see \cite{Modesto:2014lga} for more details).

In the Weyl basis  presented here the FLRW metric for conformal matter ($T_{\rm matter}\equiv T_{\rm matter}{}^\mu{}_\mu= 0$) solves exactly the non-local EOM. The Big-Bang singularity may show up in some solutions of our 
 finite theory of quantum gravity \cite{Li:2015bqa}. 

In this paper we are mostly interested in a particular solution, namely the $R^2$ 
self-inflating early Universe. 
Since we are looking for FLRW solutions the Weyl square or quartic terms do not give contribution to the 
background EOM, because Weyl tensor on FLRW background vanishes. Then we can concentrate on the following effective non-local $f(\Box, R)$ theory,
\begin{eqnarray}
\mathcal{L}_{\rm eff} =   \frac12 \kappa_D^{-2} \sqrt{|g|}\Big[ {\bf R} 
 + {\bf R}  \gamma_{\rm S}(\Box) {\bf R }
 \Big]  \: \text{ with }\: \gamma_{\rm S}(\Box) = - \frac{D-2}{4(D-1)} \frac{e^{H_0(\Box)} -1}{\Box} \, .
 \label{gammaSwf}
\end{eqnarray}
To have the scalar degree of freedom in the spectrum of perturbations we perform the substitution given by \eqref{gammaSwfscalaron}.
This theory is still unitary (as explained above) and super-renormalizable (eventually finite after adding killers), but the spectrum now consists of 
an extra degree of freedom: the $R^2$ scalaron. 

Moreover, if we expand the action in inverse powers of the scale $\Mc$ 
and use the asymptotically polynomial form of the form-factor given in \eqref{Tomboulis},
then the leading operators reconstruct exactly the effective $R^2$ theory,
\begin{eqnarray}
\mathcal{L}_{R^2} =   \frac12 \kappa_D^{-2} \sqrt{|g|}\Big[ R+ \frac{R^2}{6 M^2} 
 + O\left(\frac{R  \Box R}{M^2\Mc^2}\right)
 \Big]  \,  .
 \label{R2action}
\end{eqnarray}
Clearly, for $M<\Mc$ the $R^2$ theory is a good approximation of the non-local theory.
Therefore, we expect all the features of the $R^2$ inflation to be perfectly reproduced by our theory.

\section{$R^2$ inflationary spacetime in non-local gravity}

The goal in this section is to study the inflationary scenarios in the framework of action (\ref{TWeyl2}). Such scenarios are described by the FLRW metric, which is conformally flat (moreover for the sake of simplicity we here focus on spatially flat solutions only), hence the Weyl tensor is identically zero on the background. This implies that the Weyl-dependent terms in the action (\ref{TWeyl2}) do not contribute to the background EOM at all. However, the ${\bf C}\gamma_{\rm C} {\bf C}$ term contributes to the quadratic variation of the action. This variation is crucial and will be analyzed later on in order to study the perturbations. The ${ V_g}({\bf C})$ term does not influence the subsequent analysis entirely.
Therefore the action in question is
\begin{eqnarray}
S  = \int d^4x\sqrt{|g|}\left[ \frac{M_P^2}{2}{R}
	     + \frac\lambda 2{R}  \Fc(\Box) { R }
	     + \frac\lambda 2C_{\mu\nu\rho\sigma}\Fc_{\rm C}(\Box)C^{\mu\nu\rho\sigma}
	     -\Lambda_{\rm cc}
	        \right]\, .
		 \label{TWeyl2q}
\end{eqnarray}
Here we adopt the notations for non-local operators utilized in \cite{Craps:2014wga} in order to make use of and comparison with the previous results simpler. The identification with the present notations is rather straightforward, namely 
\begin{eqnarray}
\gamma_{\rm S}=\frac\lambda{M_P^2}\Fc(\Box) \, , \quad
\gamma_{\rm C}=\frac\lambda{M_P^2}\Fc_{\rm C}(\Box) \, .
\end{eqnarray}
$\lambda$ is a dimensionless constant which is convenient to control the scale of the $R^2$ modification.
Also the cosmological constant is (re)-introduced explicitly for generality and the term with it in the action is denoted by $-\Lambda_{\rm cc}$.

The EOM for the latter action were found already in \cite{Koshelev:2013lfm} and read as follows, 
\begin{eqnarray}
	E^\mu_\nu &\equiv& -(M_P^2+2\lambda\Fc(\Box)R)G^\mu_\nu-\Lambda_{\rm cc} \delta^\mu_{\nu}-\nonumber
	\frac12\lambda
	 {\delta}^\mu_\nu R\Fc(\Box)
	 R+2\lambda(\nabla^\mu\partial_\nu-\delta^\mu_{\nu}
	 \Box) \Fc(\Box) R
	 \nonumber\\
	 &+&\lambda{\Lc}^\mu_\nu
	 -\frac{\lambda}{2}\delta^\mu_{\nu}\left({\Lc}^\sigma_\sigma+\bar\Lc
	 \right)+
	 2\lambda
	 \left(R_{\alpha\beta}
	 +2\nabla_\alpha\nabla_\beta\right)
	 \Fc_{\rm C}(\Box)C_{\nu}^{\phantom{\nu}\alpha\beta\mu}
	 +O({\bf C}^2)=0 \, . 
	 \label{tEOM}
 \end{eqnarray}
 Here $G^\mu_\nu$ is the Einstein tensor and we have defined
 \begin{equation*}
	 {\Lc}^\mu_\nu=\sum_{n=1}^\infty
	 {{ f}}{}_n\sum_{l=0}^{n-1}\partial^\mu  R^{(l)}  \partial_\nu  R^{(n-l-1)} \, , \quad 
	 \bar\Lc=\sum_{n=1}
 ^\infty
 {f}{}_n\sum_{l=0}^{n-1} R^{(l)}    R^{(n-l)} \, ,
 \end{equation*}
where ${f}_n$ are coefficients of the Taylor expansion of the function
$\Fc(z)=\sum\limits_n f_nz^n$ and  $R^{(l)} \equiv \Box^l R$. We recall that function $\Fc(z)$ is analytic.

Notice that the mentioned above second variation of the action corresponds to the linear variation of the EOM. Since the Weyl tensor is zero, any term more than linear in ${\bf C}$ in EOM will vanish at the perturbative level. At the background level all (even linear) terms containing the Weyl tensor vanish.
Therefore, all the killer operators cubic or quartic in ${\bf C}$ do not take part in the analysis of linear perturbations. 

From (\ref{tEOM}) the trace equation is derived 
\begin{eqnarray} \con
	E=M_P^2R-\left[4 \Lambda_{\rm cc} \con
+6\lambda\Box\Fc(\Box) R+\lambda({\Lc} +2\bar\Lc)+O({\bf C}^2)\right]=0 \, .
	\label{tEOMtrace}
\end{eqnarray}
It does not contain the linear Weyl tensor contribution as the Weyl tensor is absolutely traceless.

The simplifying ansatz proposed in \cite{Biswas:2005qr} 
\begin{equation}
	\Box R=r_1R+r_2  
	\label{ansatz}
\end{equation}
really simplifies the EOM considerably.
Indeed, it turns out that
\begin{eqnarray}  
&& 
\Box^n R=r_1^n(R+r_2/r_1) \: \text{ for } \: n>0   \, , \nonumber \\
&&
\Fc(\Box)R=\Fc_1R+\Fc_2 \, , \quad   \Fc_1=\Fc(r_1) \, , \quad \Fc_2=\frac{r_2}{r_1}(\Fc_1-f_0) \, .\nonumber
\end{eqnarray}
Upon substitution of these latter relations into the EOM (\ref{tEOM}), some further algebra brings to the result obtained in \cite{Biswas:2005qr}, namely, a solution of the ansatz (\ref{ansatz}) is also a solution of 
the full non-local EOM (\ref{tEOM}) provided the following algebraic conditions are satisfied, 
\begin{eqnarray}
	\Fc^{(1)}(r_1)=0 \, , \quad \Fc_2=-\frac{M_P^2}{2\lambda}+3r_1\Fc_1 \, , \quad 4r_1\Lambda_{\rm cc}=-r_2 M_P^2 \, .
\label{relnl}
\end{eqnarray}
Here $\Fc^{(1)}(r_1)$ is the first derivative with respect to an argument. Technically, the above conditions can be re-written in a number of equivalent forms using the definitions of $\Fc_1$ and $\Fc_2$, but one has to be careful about possible divisions by ``zero'' that may arise for certain values of parameters.

We additionally stress here that presence of a cosmological term forces to have a non-zero $r_2$ while zero cosmological term forces $r_2=0$. Moreover, a reverse statement is also true. Given a solution to (\ref{ansatz}) such that $r_2\neq0$ we must have a non-zero cosmological term in the theory. Zero $r_2$ forces that no cosmological term is present.

	Moreover, we emphasize that, of course, the trace equation (\ref{tEOMtrace}) does not exhaust the whole system of gravitational field equations and, for a FLRW background, we have to consider the $(00)$-component of Einstein equations, too. Then, accounting the Bianchi identities, in addition to the terms following from the variation of action (\ref{TWeyl2q}), one can add the term $\propto a^{-4}$, i.e. energy density of ``dark radiation'', to the $(00)$-component of equations of motion.  For non-local models this aspect was discussed in details in \cite{Biswas:2010zk,Biswas:2013cha}. Note that this dark radiation may even have negative energy density. Explicitly, assuming that the solution satisfies the ansatz (\ref{ansatz}) and using relations (\ref{relnl}) one gets the following Einstein equations from (\ref{tEOM})
	\begin{equation}
		2\lambda\Fc_1\left[-(R+3r_1)G^\mu_\nu+\nabla^\mu\pd_\nu R-\delta^\mu_\nu\left(\frac {R^2}4+r_1R+\frac{r_2}4\right)\right]+{{T_r}^\mu_\nu}=0\, ,
		\label{00rad}
	\end{equation}
where $T_r$ is the stress-energy tensor of the additional radiation source. The $(00)$-component with both lower indices becomes
\begin{equation}
	2\lambda\Fc_1\left[-(R+3r_1)3H^2+\ddot R+\frac {R^2}4+r_1R+\frac{r_2}4\right]+\rho_r=0\, ,
	\label{00rad00}
\end{equation}
while some extra algebra brings this expression to a very neat form
\begin{equation}
2\lambda\Fc_1\left[\frac32 R\dot H-3H\dot R-9r_1H^2-\frac34r_2\right]+\rho_r=0 \, .
	\label{00rad00simple}
\end{equation}

	However, and this was one of the main results of \cite{Starobinsky:1980te}, the existence of inflation is incompatible with the radiation term being significant, so it should be very small both during and after inflation. It may be important before the beginning of inflation though. Nevertheless, then one 
has to consider a more general case with a non-zero spatial curvature, since the latter generates a similar term (in addition to the radiation one) with 
the large coefficient $M_P^2/M^2$ in the $(00)$-component of equations in the case of $R^2$ gravity. Thus, it is not consistent to add dark radiation to the 
model involved while still neglecting the spatial curvature. Hence, we leave this for a future work.

\subsection{Relation with local $R^2$ model}

Consider a local $R^2$ model of the form
\begin{eqnarray}
\hat S = \int d^4x\sqrt{|g|}\Big[ \frac{\hat M_P^2}{2}{R}
	     + \frac\lambda 2{R}  \hat f_0 { R }
	     -\hat \Lambda_{\rm cc}
	        \Big]  \, . 
		 \label{TWeyl2qlocal}
\end{eqnarray}
The parameters are here designated by the hats, and (\ref{TWeyl2qlocal}) is technically 
(\ref{TWeyl2q}) where the non-local operators reduced to a constant term. However, in contrary to the situation with higher powers of the $\Box$ operator we can derive (\ref{ansatz}) as one of the equations of motion for (\ref{TWeyl2qlocal})
(indeed, the trace). 
Moreover, conditions (\ref{relnl}) are simplified to
\begin{eqnarray}
\frac{\hat M_P^2}{2\lambda}=3r_1\hat f_0 \, , \quad 4r_1\hat \Lambda_{\rm cc}=-r_2\hat M_P^2 \, . 
\label{relR2}
\end{eqnarray}
Indeed, the non-local operator in between the Ricci scalars is just a constant and its derivative is always zero, moreover, 
$\Fc_2$ is identically zero here thanks to the same argument.

Comparing (\ref{relnl}) and (\ref{relR2}) we arrive to the conclusion \cite{Koshelev:2014voa} that any (conformally flat) solution of a local $R^2$ model (\ref{TWeyl2qlocal}) is a solution of the EOM coming from (\ref{TWeyl2q}) with the following identification of parameters, 
\begin{equation}
	\hat M_P^2=M_P^2+2\lambda \Fc_2,\quad\hat\Lambda_{\rm cc}=\Lambda_{\rm cc}\frac{\hat M_P^2}{M_P^2} \,  .
	\label{R2nlR2}
\end{equation}
Technically, $M_P^2$ or $\hat M_P^2$, or both can turn out to be negative. 
Their identification requires $\Fc_2=0$, provided we want the same Planck masses in both theories. Additionally a radiation source may be needed in either local or non-local model to support a particular solution to equations of motion. Density of radiation in either case can be computed using (\ref{00rad00simple}) while the sign of the radiation energy density is ``solution dependent".

We now understand that even though the non-local model with the ansatz (\ref{ansatz}) looks pretty much the same as a local $R^2$ theory, they are different in a number of parameters left. Therefore, we still have more freedom of adjustment in the non-local model even upon the ansatz (\ref{ansatz}) imposition.

\subsection{$R^2$ inflation}\label{sec32}

The $R^2$ inflationary model is given by a local $R^2$ gravitational theory of the type (\ref{TWeyl2qlocal}) without a cosmological term at all, i.e. $\hat\Lambda_{\rm cc}=0$. Relations (\ref{relnl}) imply that $r_2=0$ and the ansatz reduces to
\begin{equation}
	\Box R=r_1 R\, ,
	\label{ansatzr20}
\end{equation}
which is exactly the trace equation in $R^2$ gravity.
An analytic solution to the above equation cannot be obtained in full. However, the vital piece of information for us is that $R^2$ inflation is indeed a solution of a local $R^2$ gravity model. As perhaps the most crucial consequence supported by the previous subsection is that the same solution is a solution in our non-local model upon adjustment of the parameters dictated by (\ref{relnl}).

In this paper we are mostly aimed at studying the nearly dS phase of $R^2$ inflation. This regime is given by the following expressions:
\begin{eqnarray}
 a(t) &\approx& a_0 ({t_s - t})^{-1/6} \, e^{-r_1 (t_s -t)^2/12} \, ,
	\label{starinffactor} \\
	H=\frac{\dot a}a&=&  \frac{{r_1}(t_s-t)}{6}+\frac{1}{6(t_s-t)}+ ...\, ,\label{starH}\\
	R = 6(\dot H +2H^2) &=&  \frac{r_1^2(t_s-t)^2}{3}- \frac{r_1}{3} +\frac{4}{3(t_s-t)^2} + ...\, ,
\label{starR}
\end{eqnarray}
where $t_s$ corresponds to the end of inflation when $|\dot H| \sim H^2$, dot means the derivative with respect to the cosmic time $t$, and all
formulae are valid for $\sqrt{r_1}(t_s-t)\gg 1$. Note that we must have $r_1 >0$ for inflation to be metastable. In this case, the Universe undergoes inflation when $t <t_s$, and for $t>t_s$, it has a graceful exit to the subsequent power-law expansion stage with $a(t) \propto t^{2/3}$ modulated by small oscillations.\footnote{It was shown in \cite{Craps:2014wga} that the scale factor in (\ref{starinffactor}) gives an \textit{exact} solution to the full system of non-local (or local as well) equations of motion provided we introduce a small negative cosmological term. The presence of a cosmological term in a local $R^2$ model describing inflation is questionable as this introduces an unnecessary extra parameter.} In principle, $t_s-t$ can be arbitrary large in these formulae, so the local $R^2$ inflation never approaches the exact dS phase with a constant $R$.

Note that to obtain the power-law multiplier in. (\ref{starinffactor}) and the last terms in the other two equations, one has to go beyond the
leading order in the slow-roll approximation, see  the redivation in the Appendix~\ref{apstarobinsky}. For this reason, the power-law exponent is different from that in 
Eq. (7.26) in \cite{Mukhanov:1990me} where this correction was not taken into account.

In order to see what happens to a possible radiation source we analyze equation (\ref{00rad00simple}) accounting that $r_2=0$ as well as identifying $r_1=M^2$ and $\lambda\Fc_1=M_P^2/(6M^2)$ where $M$ is the scalaron mass. All this together yields
\begin{equation}
	\frac{M_P^2}{3M^2}\left[\frac32 R\dot H-3H\dot R-9M^2H^2\right]+\rho_r=0 \, .
	\label{00radlocal}
\end{equation}
One can see that a simple substitution of expressions (\ref{starH}) and (\ref{starR}) up to the leading order (i.e. keeping only the first terms) yields $\rho_r=0$ which is in the total agreement with the physical arguments presented above right after equation (\ref{00rad00simple}).

\subsection{Non-local $R^2$ inflation}

To embed $R^2$ inflation as it is in a local model in our non-local framework we essentially should satisfy relations (\ref{relnl}) when we already assume $\Lambda_{\rm cc}=r_2=0$. We therefore focus on parameters and non-local operators.
Combining (\ref{gammaSwf}) and (\ref{gammaSwfscalaron}) we arrive to the following wishful form-factor in $D=4$, 
\begin{eqnarray}
&& \frac{\lambda}{M_P^2} \mathcal{F}(\Box) =  - \frac{1}{6 \Box} \left[  e^{H_0(\Box)}\left( 1 - \frac{\Box}{M^2} \right)   - 1 \right] \, . 
     \label{H2H1}
\end{eqnarray}
Evaluating the second relation in (\ref{relnl}) one, after canceling all manifestly non-zero factors on opposite sides, gets:
\begin{eqnarray}
\frac{r_2}{r_1}(1-e^{-H_0(r_1)}) \left(1-\frac{r_1}{M^2}\right) = 3r_1\left(1-\frac{r_1}{M^2}\right) \, . 
\label{H2H1var}
\end{eqnarray}
In the case $r_1\neq M^2$ one can cancel the corresponding factors, but as a consequence one gets with necessity $r_2\neq0$. The latter implies a non-zero cosmological constant due to the last relation in (\ref{relnl}) and contradicts our initial assumptions. Notice that this outcome appears irrespectively of the particular solution solely as the result of our choice of the form-factor.

The absence of a cosmological constant term, given the form-factor (\ref{H2H1}), is only compatible with $r_1=M^2$. Notice that this value for the parameter $r_1$ is exactly like in a local $R^2$ gravity theory.
The remaining question is to maintain the first relation in (\ref{relnl}). Starting from (\ref{H2H1}) and assuming $r_1=M^2$ we arrive at
\begin{equation}
	H_0(r_1)=0\, .\label{H0zeroatr1}
\end{equation}
Even though the condition looks extremely simple it is incompatible with a proposed above form for $H_0(z)$ given by (\ref{Tomboulis}). Indeed, such form produces $H_0(z)$ positive for any non-zero argument and $H_0(0)=0$. In our case $r_1$ should not be trivial and therefore we are set to slightly modify the form of entire function $H_0(z)$. 
A natural and very simple form-factor compatible with $H_0(r_1) = 0$ reads as follows,\footnote{
	Notice that even if $\Lambda_{\rm cc}\neq 0$ and consequently $r_2\neq 0$ an assumption $r_1\neq M^2$ leads to an impossible (in real numbers) relation $e^{-H_0(r_1)}=-\frac54<0$. So, we again must require (\ref{H0zeroatr1}) and therefore implement changes to the form-factor $H_0$.}
\begin{eqnarray}
H_0(\Box)=\frac{a}{2} \left[ \Gamma \left(0, p_\gamma(\Box)^2 \right)+\gamma_E  + \log \left( p_\gamma(\Box)^2 \right) \right] \, , \quad p_\gamma (\Box) = {\Box^{\gamma-1} (\Box - M^2)} \, .
\label{StaroTomboulis}
\end{eqnarray}
We here remind that in $D=4$ we achieve super-renormalizability with only one-loop divergences for the minimal choice 
$\gamma = 3$, therefore for this particular value of $\gamma$ we get:
\begin{eqnarray}
 \quad p_3 (\Box) =  {\Box^{2} (\Box - M^2)} \quad \Longrightarrow \quad e^{H_0(p_3(\Box))}\Big|_{\Box = 0} = e^{H_0(p_3(\Box))}\Big|_{\Box = M^2} = 1\, .  
\end{eqnarray}

	Analyzing a possible radiation source we look at equation (\ref{00rad00simple}) accounting that $r_2=0$. Moreover, the above construction of the operator function $\Fc(\Box)$ (see (\ref{H2H1})) and especially conditions $r_1=M^2$ and (\ref{H0zeroatr1}) imply that $\lambda\Fc_1 \equiv \lambda\Fc(r_1)=M_P^2/(6M^2)$. This is exactly the coupling in front of the $R^2$ term in a local inflationary model. Combining all of this together one yields
\begin{equation}
	\frac{M_P^2}{3M^2}\left[\frac32 R\dot H-3H\dot R-9M^2H^2\right]+\rho_r=0 \, ,
	\label{00radnonlocal}
\end{equation}
which is an equation identical to the local model case (\ref{00radlocal}).
This means that for a dense open subset of all solutions of the local
$R^2$ model having a sufficiently long slow-roll inflationary stage (\ref{starinffactor}-\ref{starR}) required by observational data, the radiation term $\rho_r$ becomes unimportant once inflation begins (and even after its end) in our non-local model as well.

%%%%%%%%%%%%%%%%%%%%%%%%%%%%%%%%%%%%%%%%%%%%%%%%%%%%%%%%%%%%%%%%%%%%%%%%%%%%%%%%%%%%%%%%%%%%%
\section{Finite quantum gravity with three propagating degrees of freedom}
%%%
Let us here expand about finiteness of the theory for what concerns the Newton and cosmological constant couplings. In section (\ref{FiniteS}) we reviewed a special class of finite theories with monomial UV behaviour of the form factors \cite{Modesto:2014lga}.  However, for the UV completion of the local $R^2$ model the form factor $\mathcal{F}$ is polynomial, and not monomial, in the UV regime. In particular, the next to the leading order  term in the UV behaviour  in the large momenta limit is forced to depend on the scalaron mass $M$. Or in other words the effective mass of the scalaron is determined from the next to leading higher derivative term in the UV regime. We would like to remind that in this theory we have two propagating degrees of freedom in the metric fluctuations and one in the scalaron field.

Focusing on the structure of the theory in UV we can have non-zero beta functions for the Newton constant 
($\beta_{G_{\rm N}}$) and the cosmological constant ($\beta_{\bar{\lambda}}$). Moreover if we treat the scalaron mass as the fundamental coupling the integration of the RG equations for the two couplings in front of the two quadratic in curvature operators is non-trivial. Namely, the beta functions depend on the coupling itself, which is actually the mass of the scalaron. We can view the system of beta functions from the other perspective too and it still can be integrated quite easily. The mass of the scalaron may be considered not as  a fundamental coupling, it is instead read from  $\omega_{\gamma-1}$ or $\omega_{\gamma-2}$. The last parameters do not run under change of energy scale. The theory is like any higher derivative theory, where beta functions are plain. Only specific relation between  them define the scalaron mass $M$. The only issue is how to have non-vanishing mass of the scalaron but at the same time vanishing of the beta functions for $G_N$ and cosmological constant. It is obvious since divergences are local that divergences (and hence beta functions) of the theory do not depend on the curvature of background and we easily read them as from the theory on flat spacetime background (in the short distance limit any non-singular spacetime is effectively flat). Below we show how to obtain this for a specific choice of the form factor.

%Nevertheless, an extra operator can secure vanishing of such beta function $\beta_{G_{\rm N}}$, namely 
%\be
%s^{(3)}_{\rm C} C_{\mu\nu\rho \sigma} \Box^{\gamma - 2} C^{\mu\nu}_{\delta \tau} C^{\rho \sigma \delta \tau} \, . 
%\ee
%It is easy to see using the background field method that there is a contribution to $\beta_{G_{\rm N}}$
%linear in $s^{(3)}_{\rm C}$. 
%Also the beta function for the cosmological constant is in general non-zero, but the following choice of the polynomial appearing in the entire function $H_2$, which appears in the form factor 
%$\mathcal{F}_{\rm C}$, can do the job of making the theory finite,
%\be
%p_{\gamma}(\Box) = \Box^{\gamma+1} + a_{\gamma-2} \Box^{\gamma-1} \, . 
%\ee
%Indeed, $a_{\gamma-2}$ contributes linearly to the beta function of the cosmological constant \cite{Asorey:1996hz} and we can always select it out to make zero the beta function. 

%Moreover, there is another much simpler 
There is a simple way to make harmless the UV non-monomial contribution resulting in the limit $\Box \rightarrow +\infty$ of the  form factor $\mathcal{F}$. We can just replace the polynomial (\ref{StaroTomboulis}) with the following one,
\be
p_\gamma (\Box)= 
\frac{M^2}{\mathcal{M}^2} \frac{1}{\mathcal{M}^{2 \gamma } } \Box^{\gamma - 5} (\Box^2 + M^2 \Box + M^4)^2 (\Box - M^2)\, ,  \quad \gamma>5 \,  .
\label{poly2}
\ee
The asymptotic behaviour of $\mathcal{F}$ now reads
\be
%\hspace{-0.5cm}
&&\hspace{-0.5cm}
 \mathcal{F} (\Box) \,\, \rightarrow \,\, 
\frac{M_P^2}{\lambda} \, \frac{M^2}{\mathcal{M}^{2 \gamma +2}} \Box^{\gamma-5}(\Box^2 + M^2 \Box + M^4)^2 (\Box - M^2)\frac{(\Box -M^2)}{6 M^2 \Box} \nonumber \\
&&  \hspace{-0.5cm}
= \frac{M_P^2}{\lambda}  \, \frac{1}{6 \mathcal{M}^{2 \gamma+2} }\Box^{\gamma-6}(\Box^3 - M^6)^2 = \frac{M_P^2}{\lambda}  \, \frac{1}{6 \mathcal{M}^{2 \gamma+2} }
(\Box^{\gamma} -2 \Box^{\gamma-3} M^6 + \Box^{\gamma-6} M^{12} )  .  
\label{polyF2}
\ee
Following the notation of reference \cite{Asorey:1996hz}, now the coefficients $\omega_{\gamma -1}$ and $\omega_{\gamma-2}$ are identically zero, which means there are no operators quadratic in gravitational curvature with in between neither $\Box^{\gamma-1}$ nor $\Box^{\gamma-2}$. Then the beta functions for $G_N$ and the cosmological constant are identically zero. Moreover, the coefficient 
$\omega_\gamma$ (we again refer to the paper \cite{Asorey:1996hz}) is independent on the scalaron mass $M$. Therefore, the beta functions for $R^2$ and ${\bf C}^2$ can only depend on the scale 
$\mathcal{M}$ \con and the coefficients $s_{\rm C}^{(1)}$ and $s_{\rm C}^{(2)}$, the latter two can be selected to make zero the beta functions for the operators quadratic in the curvature. It is crucial that the coefficients $\omega_{\gamma}$, $s_{\rm C}^{(1)}$ and $s_{\rm C}^{(2)}$ do not get any divergent renormalization, or, which is the same, that $\mathcal{M}$ is not renormalized. Indeed, the theory is super-renormalizable and such coefficients appear in front of higher derivative operators. 
Note that the polynomial (\ref{poly2}) takes zero value in 
$\Box = 0$ and in $\Box = M^2$ as it is required by the classical solution (see the previous section).

If the theory is only super-renormalizable we expect perturbative logarithmic corrections to the operators $R^2$ and ${\bf Ric}^2$ (for the sake of simplicity we here consider the case $\beta_{G_{\rm N}}=\beta_{\bar{\lambda}}=0$), namely in the UV 
\be 
R \log \frac{ - \Box}{\mu^2} R \quad {\rm and} \quad {\bf Ric} \log \frac{ - \Box}{\mu^2} {\bf Ric}\, , 
\ee
where $\mu$ is a general renormalization scale. 
%which, however, do not move the scalaron pole because the arbitrariness of the finite piece is reflected in the explicit  appearance of an arbitrary mass scale $\mu$. 
We can write the one-loop dressed propagator in momentum space as follows \cite{Briscese:2013lna},
\be
&& \hspace{-0.5cm}
 \mathcal{O}^{-1} = - \frac{P^{(2)} e^{-H_2(p^2)}}{ p^2 \Pi_2(p^2)} + \frac{ P^{(0)} e^{- H_0(p^2) }  }{ (D-2) p^2 \Pi_0(p^2)} \label{dressed}
 \\
&& \hspace{-0.5cm}
= - 
\frac{P^{(2)} e^{-H_2(p^2)}}{ p^2  \left[1+ e^{- H_2 }    p^2 c_0  \log \left(  \frac{p^2}{\mu^2} \right) 
%- i \pi \theta(-k^2) \right) 
\right] } %\nonumber  \\&&
  + \frac{ P^{(0)} e^{- H_0(p^2) }  }{ (D-2) p^2 \left[ \left( 1 + \frac{p^2}{M^2} \right)+ e^{- H_0(p^2) } p^2 \bar{c}_0    \log \left(  \frac{p^2}{\mu^2} \right) % - i \pi \theta(-k^2) \right)
   \right]}\, . 
  \nonumber 
\ee
The constant $c_0$ and $\bar{c}_{0}$ have both inverse mass square dimension and are related to the beta functions for the counterterms quadratic in the curvature. The dressed propagator (\ref{dressed}) develops an infinite number of complex conjugate poles in both sectors: the spin two and the spin zero.
Therefore, the scalaron pole survives at quantum level iff $\bar{c}_{0}=0$, which means that 
at least the beta function for the counterterm $R^2$ must vanish. This is consistent with the finiteness
of the theory at quantum level. 

Now we apply the most popular subtraction point at $p^2 = -M^2$, where the scalaron is on the mass shell. 
This means we want the finite part of $\Pi_0(p)$ to vanish when $p^2 = -M^2$. 
Therefore, the finite part is uniquely specified by defining 
\be
\Pi_{0, \rm R}(p^2) = \Pi_0(p^2) - \Pi_0(-M^2).
\ee
Finally, the one-loop propagator reads
\be
\hspace{-0.1cm}
- \frac{P^{(2)} e^{-H_2(p^2)}}{ p^2
 \left[1+ c_0 \, e^{- H_2(p^2) }    p^2   \log \left( - \frac{p^2}{M^2} \right)  \right] }
  + \frac{ P^{(0)} e^{- H_0(p^2) }  }{ (D-2) p^2
\left[ \left( 1 + \frac{p^2}{M^2} \right) + \bar{c}_0 \, 
e^{- H_0(p^2) } 
p^2    \log \left( -\frac{p^2}{M^2} \right) \right]}. \nonumber 
\ee
This is of course just a particular identification of the renormalization scale $\mu$ that consistently preserves the mass spectrum of the theory without introducing mixing between $\mathcal{M}$ and $M$.

Notice that at perturbative level there is no hierarchy problem for the finite theory because there are no divergences forcing us to renormalize $M_P^2  = 1/8 \pi G_N$ and/or the scalaron mass $M$. Both the scales can coexist in a finite theory of gravity without perturbative mixing (see the discussion above). 
Moreover, the finite contributions in the ultraviolet regime look like:  
%\be
%R(1/\Box^n) \quad {\rm or}  \quad {\bf C} (1/\Box^n) {\bf C} \quad (n>0), 
$R(1/\Box^n) R$ or ${\bf C} (1/\Box^n) {\bf C}$ ($n>0$), 
%\ee
and they can not significantly move the poles. 

It deserves to be noticed that both the finite and the super-renormalizable theory with polynomial behaviour 
(\ref{poly2}) are perturbatively consistent with zero value (or any small value) of the cosmological constant at quantum level.

In addition to these theoretical arguments, phenomenology of inflation, namely the specific form of deviation of the power spectrum of primordial scalar (density) perturbations measured in \cite{Ade:2015xua,Ade:2015lrj,Array:2015xqh} from the exactly scale-invariant (or, Harrison-Zeldovich) one, suggests that gravity 
is approximately scale-free in the range of Ricci curvatures between $M^2$ and ${\cal M}^2$. So, in the local limit, terms containing higher 
order powers of the Riemann curvature are somehow suppressed relative to quadratic ones for a large range of curvatures. That is why it is interesting and important to study in the first approximation what happens if they are absent at all.

Moreover, in our theory the higher in curvature Weyl  killer operators in (\ref{WeylKiller}) do not influence the two-point function around the de Sitter space at the classical level because they are constructed out of the Weyl tensor, which vanishes on the background. At quantum loop-level we expect such operators to give perturbative contributions only to finite terms in the UV-finite complete theory. These last contributions will be perturbatively suppressed in the coupling constant.

\section{Cosmological perturbations during inflation}%\label{sec_lin_EOM}

A vast amount of a related work in non-local gravities was already done in 
\cite{Biswas:2012bp,Biswas:2016etb} and \cite{Craps:2014wga}. 
%In those papers the model (\ref{TWeyl2q}) without the piece containing the Weyl tensor was considered. 
In the paper \cite{Biswas:2016etb} the second variation of the most general action including the Weyl tensor was derived using the spin decomposition of the gravitational fluctuation that we will review and use in section \ref{pippo2}. However, in section \ref{pippo1} we present a general self-adjoint second order variation of the nonlocal Weyl squared term without using any decomposition of the fluctuations (see formula (\ref{hmunugeneric})). The general variation (\ref{hmunugeneric}) is essential in the evaluation of the beta functions for the $G_N$ coupling, but it is also useful to obtain the combination of beta functions for the couplings in front of the $R^2$ and ${\rm \bf Ric}^2$ terms. Moreover, we also need (\ref{hmunugeneric}) to compute the exact graviton propagator around any maximally symmetric background (which is either of dS, anti-dS or Minkowski)\footnote{After the present paper was submitted to arxiv and to the journal, the paper \cite{Biswas:2016egy} co-authored by one of the present authors appeared in which the detailed computation of the second order variation of non-local quadratic in curvature actions around maximally symmetric background using the spin decomposition of the metric fluctuation was presented.}.
Here we breakdown the essence of the previous results complementing it with new developments related to the Weyl tensor term. Moreover, we gain certain simplifications considering a model without an explicit cosmological term.

To find out the parameters of inflation such as power spectra, spectral tilts and the tensor to scalar power spectra ratio we have to quantize perturbations in the inflationary background. To accomplish this task we need the second order variation of the action around the inflationary (in case of $R^2$ inflation in fact a nearly dS) background. However, to perform certain mathematical manipulations we have to have the linear variation of equations of motion. Since the latter is even simpler to derive we will elaborate on both: linear variation of the EOM and the quadratic variation of the action.

We start by introducing the following basic notation, 
\begin{equation}
	g_{\mu\nu}={\bar g}_{\mu\nu}+h_{\mu\nu} \, . 
	\label{hmunu4}
\end{equation}
Hereafter the bar is used to designate the background values. We shall proceed according to the following plan:
\begin{itemize}
	\item First, we will derive the second variation of the action around a nearly dS background in terms of covariant metric perturbation $h_{\mu\nu}$. 
	\item Second, we will analyze the spin-2 mode, i.e. transverse and traceless part of $h_{\mu\nu}$. This will lead to an important constraint on the form-factor $\Fc_{\rm C}$. 
	\item Third, we will introduce the canonical, in cosmology, ADM formalism and we will perform the study of tensor perturbations around the dS background. 
	\item Fourth, we will derive the linearized equations for scalar perturbations and quantize the scalar perturbations around a nearly dS background. 
\end{itemize}
In the last subsection we will compute the ratio of power spectra of tensor and scalar perturbations at the end of inflation in our non-local theory.

\subsection{Second order variation of action (\ref{TWeyl2q}) in terms of $h_{\mu\nu}$}
\label{pippo1}

Considering Einstein-Hilbert and cosmological (just for the generality) terms in (\ref{TWeyl2q}) one can derive the following second order variation
\begin{eqnarray}
\delta^2 S_{{\rm EH} +\Lambda_{\rm cc}}=\int
d^4x\sqrt{|\bar g|}\frac{M_P^2}2 \left[\delta_{\rm EH}-\frac2{M_P^2}\Lambda_{\rm cc}\delta_g\right] \, , 
\end{eqnarray}
\label{deltaEHfull}
where we have introduced the following definitions, 
\begin{eqnarray}
	\delta_{\rm EH}&=& \left(\frac14 h_{\mu\nu}\bar\Box
h^{\mu\nu}-\frac14h\bar \Box h+\frac12h\bar\nabla_\mu\bar\nabla_\rho
h^{\mu\rho}+\frac12\bar\nabla_\mu h^{\mu\rho}\bar\nabla_\nu h^\nu_\rho\right) \nonumber
\\
&+&(hh^{\mu\nu}-2h^\mu_\sigma h^{\sigma\nu})\left(\frac18
\bar g_{\mu\nu}\bar R-\frac12 \bar R_{\mu\nu}\right)-\left(\frac12\bar R_{\sigma\nu}h^\sigma_\rho
h^{\nu\rho}+\frac12\bar R^\sigma_{\rho\nu\mu}h^\mu_\sigma h^{\nu\rho}\right) \, , \\
 \delta_g&=& \frac{h^2}{8}-\frac{h_{\mu\nu}^2}{4}\, .
\label{delta0lambda}
\end{eqnarray}
One can verify this result against, e.g. \cite{Christensen:1979iy}. This is a generic fully covariant variation around any background.  

One important remark must be made here. Below we are greatly focused on considering the nearly dS phase of the evolution of the $R^2$ inflation. It is crucial that this phase happens due to the presence of the $R^2$ term and not due to a cosmological term, since the latter is absent at all in our model of inflation. This implies that we do not have the usual in GR relation between $\Lambda_{\rm cc}$ and the background scalar curvature $\bar R$. We nevertheless are allowed to consider the Riemann tensor as having a nearly dS form.

Next we vary the term quadratic in the scalar curvature. Doing this around a nearly dS spacetime\footnote{In fact this is the only term that for the moment we can analyze around a nearly dS background only.} we can show that the terms related to the variation of d'Alembertians inside the function $\Fc$ cancel out. The fact that $\bar R$ is almost constant in this approximation helps a lot. We thus obtain the following variation (this can be checked against \cite{Craps:2014wga,Biswas:2016etb}), 
\begin{equation}
	\delta^2 S_{R^2}=\int d^4x\sqrt{|\bar g|}  \frac{\lambda}{2}\left[(2\delta_{\rm EH}-\bar R\delta_g)f_0\bar R+\delta R  \Fc(\bar\Box)\delta R\right] \, , 
	\label{d2RFRfullds}
\end{equation}
where the variation of the scalar curvature reads,
\begin{equation}
	\delta R=(-\bar R_{\mu\nu}+\nabla_\mu\nabla_\nu-g_{\mu\nu}\bar\Box)h^{\mu\nu} \, .
	\label{deltaRfull}
\end{equation}
Note that the expression for $\delta R$ is valid for any background.

The last piece we need is the second variation of the Weyl-dependent term. It turns out that this term is quite simple as long as we consider conformally flat backgrounds (like FLRW in general and the $R^2$ inflationary solution in particular). Indeed, the Weyl tensor is zero on such backgrounds and therefore the only non-trivial combination which persists at the second order in the gravitational fluctuation reads as follows, 
\begin{equation}
	\delta^2 S_{C^2}=\int d^4x\sqrt{|\bar g|}\frac{\lambda}{2}(\delta C_{\mu\nu\alpha\beta})\Fc_{\rm C}(\bar \Box)(\delta C^{\mu\nu\alpha\beta}) \, . 
	\label{d2CFCfullds}
\end{equation}
We stress that this formula is valid around any conformally flat background. 
In the next sections we will present this formula for scalar perturbations in the ADM formalism and for an arbitrary scale factor. However, its expression in terms of $h_{\mu\nu}$ for the moment could be derived only around a nearly dS background (strictly speaking around a nearly maximally symmetric background). The corresponding expression reads as follows, 
\begin{eqnarray}
\frac{1}{2}\delta^{2} \!\! &\bigg(&\int d^{4}x\sqrt{|g|}C_{\mu\nu\rho\sigma}{\cal F}_{{\rm C}}\left(\bar\Box\right)C^{\mu\nu\rho\sigma}\bigg)=  \nonumber \\
&=& \int d^{4}x\sqrt{|\bar g|}\left\{h_{\mu\nu}\left(\frac{\bar{R}^{2}}{36}-\frac{\bar{R}}{4}\bar{\Box}+\frac{1}{2}\bar{\Box}^{2}\right){\cal F}_{\rm C}\left(\bar{\Box}+\frac{\bar{R}}{3}\right)h^{\mu\nu}\right.\nonumber\\
&-&h\left[\left(\frac{\bar{R}^{2}}{144}-\frac{\bar{R}}{16}\bar{\Box}+\frac{1}{8}\bar{\Box}^{2}\right){\cal F}_{{\rm C}}\left(\bar{\Box}+\frac{\bar{R}}{3}\right)+\left(\frac{\bar{R}}{48}\bar{\Box}+\frac{1}{24}\bar{\Box}^{2}\right){\cal F}_{{\rm C}}\left(\bar{\Box}+\bar{R}\right)\right]h\nonumber\\
&-&h_{\mu\nu}\bar{\nabla}^{\mu}\left(\frac{\bar{R}}{12}+\bar{\Box}\right){\cal F}_{{\rm C}}\left(\bar{\Box}+\frac{3}{4}\bar{R}\right)\bar{\nabla}^{\rho}h^{\nu}{}_{\rho}%\nonumber\\
+\frac{1}{3}h_{\mu\nu}\bar{\nabla}^{\mu}\bar{\nabla}^{\nu}{\cal F}_{{\rm C}}\left(\bar{\Box}+\bar{R}\right)\bar{\nabla}^{\rho}\bar{\nabla}^{\sigma}h_{\rho\sigma}\nonumber\\
&+&\left.h\left(\frac{\bar{R}}{6}+\frac{1}{3}\bar{\Box}\right){\cal F}_{{\rm C}}\left(\bar{\Box}+\bar{R}\right)\bar{\nabla}^{\mu}\bar{\nabla}^{\nu}h_{\mu\nu}\right\}\, .\label{hmunugeneric}
\end{eqnarray}

\subsection{Covariant spin-2 excitations and restrictions on form-factor $\Fc_{\rm C}(\Box)$}
\label{pippo2}

The covariant spin-2 excitation of the generic metric variation $h_{\mu\nu}$ is the transverse and traceless part which we denote as $h^\perp_{\mu\nu}$. Since it enjoys the conditions 
$\bar \nabla^\mu h^\perp_{\mu\nu}=\bar g^{\mu\nu}h^\perp_{\mu\nu}=0$, its substitution to the general second variation is rather simple (moreover, once can use the results of \cite{Biswas:2016etb} in order to check the subsequent expression). As long as we use formula (\ref{hmunugeneric}) the subsequent analysis in this subsection is valid only around a nearly dS background. Performing some algebra one gets 
\begin{equation}
	\begin{split}
		\delta^2  S_{\perp}=\int d^4x\sqrt{|\bar g|}\left\{\frac{M_P^2}2\right.&\left[\frac14h^\perp_{\mu\nu}\left(\bar\Box-\frac{2\bar R}3\right){h^\perp}^{\mu\nu}+\frac2{M_P^2}\Lambda_{\rm cc}\frac14{h^\perp_{\mu\nu}}^2\right]\\
		+\frac\lambda2&\left[2\frac14h^\perp_{\mu\nu}\left( \bar\Box-\frac{2\bar R}{3} \right){h^\perp}^{\mu\nu}+\bar R\frac14{h^\perp_{\mu\nu}}^2\right]f_0\bar R\\
		+\frac\lambda2&\left.2\frac14h^\perp_{\mu\nu}\left(\bar\Box-\frac{\bar R}6\right)\left(\bar\Box-\frac{\bar R}3\right)\Fc_{\rm C}\left(\bar\Box+\frac{\bar R}3\right){h^\perp}^{\mu\nu}\right\} \, .
	\end{split}
	\label{d2spin2ds}
\end{equation}
Here we keep some factors not cancelled and some similar terms not grouped in order to make the computation track transparent.

We recall that the relations (\ref{relnl}) must hold for the solution.
\begin{equation*}
	\Fc_2=0 \, , \quad f_0=\Fc_1 \, , \quad 2f_0\bar R+\frac{M_P^2}\lambda=2\Fc_1(\bar R+3r_1) \, .
\end{equation*}
Moreover, in the nearly dS phase one has 
\begin{equation}
	r_1\ll\bar R\,.
\end{equation}
The latter comparison follows from (\ref{starH}) and the discussion thereafter.

Under all these approximations we can neglect the first line in formula (\ref{d2spin2ds}) while the rest neatly combines in the following variation, 
\begin{equation}
	\delta^2 S_{\perp , {\rm inf}}=\int d^4x\sqrt{|\bar g|}\frac\lambda4h^\perp_{\mu\nu}\left(\bar\Box-\frac{\bar R}6\right)\left\{\Fc_1\bar R+\left(\bar\Box-\frac{\bar R}3\right)\Fc_{\rm C}\left(\bar\Box+\frac{\bar R}3\right)\right\}{h^\perp}^{\mu\nu} \, , 
	\label{d2spin2dsinf}
\end{equation}
where ``inf'' stands for inflation. Here we easily distinguish the standard dS propagator factor $\bar\Box-\bar R/6$ for the spin-2 mode, but on top of this we have a new non-local operator in the curly brackets. The concept of non-local theories stays that the quadratic form of the d'Alembertian operator defines as many degrees of freedom as many roots this form as the function of $\bar \Box$ has. 
The factor $\bar\Box-\bar R/6$  
gives the standard pole of the propagator. 
If we require that extra poles do not appear (in fact an extra pole will necessarily be a ghost), then the following function has no roots,
\begin{equation}
	\Pc(\bar\Box)=1+\frac1{\Fc_1\bar R}\left(\bar\Box-\frac{\bar R}3\right)\Fc_{\rm C}\left(\bar\Box+\frac{\bar R}3\right) \, . 
	\label{newroots}
\end{equation}
Mathematically this implies that $\Pc(\bar\Box)$ is the exponential of an entire function of $\bar\Box$, namely  
\begin{equation}
\Pc(\bar\Box)=e^{-2 \omega(\bar\Box)} \, , 
	\label{newrootsno}
\end{equation}
where $\omega(\bar\Box)$ is some entire function. We emphasize that the above derived restriction on the form-factor $\Fc_{\rm C}$ must be treated as a general requirement and should propagate to any regime in the evolution of the background because we want the theory to be always healthy. However, the value of scalar curvature used to obtain the restriction is the one during the nearly dS inflationary  phase.

An example of form-factor $\Fc_{\rm C}$ compatible with (\ref{newrootsno}) reads 
\begin{equation}
\mathcal{F}_{\rm C} (\Box) = \mathcal{F}_1  \bar{R}  \frac{e^{H_2(\Box - \frac{2}{3} \bar{R} ) } - 1}{\Box - \frac{2}{3} \bar{R} }\, .
\label{FcR}
\end{equation}
Indeed, (\ref{FcR}) generates the desired behaviour for the operator $\Pc$, namely 
\begin{equation}
\mathcal{P}(\Box) = e^{H_2(\Box - \frac{2}{3} \bar{R} ) }\, .
\end{equation}
Redefining the field $\hat h{}^\perp_{\mu\nu}=e^{-\omega(\bar\Box)}h^\perp_{\mu\nu}$ we finally get:
\begin{equation}
	\delta^2 S_{\perp , {\rm inf}} =\int d^4x\sqrt{|\bar g|}\frac\lambda4\Fc_1\bar R\hat h{}^\perp_{\mu\nu}\left(\bar\Box-\frac{\bar R}6\right){\hat h{}^\perp}^{\mu\nu} \, . 
	\label{d2spin2dsquantum}
\end{equation}
One can check that the action of the $\bar\Box$ operator does not spoil the properties of the original field to be transverse and traceless.

\subsection{Tensor perturbations}

Let us recall that the most standard formalism for studying the cosmological perturbations is the ADM decomposition of the spacetime. In what follows we proceed in the canonical way classifying the perturbations with respect to the representation of the 3-dimensional symmetry group. That is this section is about standard tensorial perturbations. In contrast to this, we used the covariant approach in the previous subsection mainly in order to derive the constraint (\ref{newrootsno}). This constraint will manifest its usefulness in the analysis of scalar perturbations. We remind that the scalars, vectors, and tensors do not mix at the linear order and can be studied independently. Moreover, we omit vectors and concentrate on cosmologically important scalars and tensors only starting with the tensors.

The line element for tensor perturbations reads 
\begin{equation}
	ds^2=a(\tau)^2\left[-d\tau^2+(\delta_{ij}+2h_{ij})dx^idx^j\right]\, ,
\label{mFrtensors}
\end{equation}
While all the notations are collected in the appendix~\ref{ap:notation}, we just recall that $\tau$ is the conformal time.
The three-dimensional metric is just the delta symbol as we already stated our wish to consider only spatially flat solutions in this paper. Tensor $h_{ij}$ is transverse and traceless and, moreover, it is gauge invariant. One can check that $h_{ij}$ forms a subset of the four-dimensional tensor 
$h^\perp_{\mu\nu}$ considered in the previous subsection. This very nice fact saves us from a lot of computations. Indeed, the last formula of the previous subsection is ready for the quantization of the tensor perturbations. Furthermore, reviewing the results of \cite{Craps:2014wga} we see that our formula (\ref{d2spin2dsquantum}) is almost exactly equation (4.23) in \cite{Craps:2014wga} modulo a normalization of the tensor field. Going through the results of \cite{Craps:2014wga} we write down the power spectrum for the tensor modes as
\begin{equation}
	|\delta_h|^2=\frac1{2\pi^2\lambda\Fc_1\bar R}\frac{k^2}{a^2}e^{2\omega(\bar R/6)} \, , 
	\label{pshexpsub}
\end{equation}
in the sub-Hubble regime and
\begin{equation}
	|\delta_h|^2=\frac{H^2}{2\pi^2\lambda\Fc_1\bar R}e^{2\omega(\bar R/6)} \,  ,
	\label{pshexpsuper}
\end{equation}
in the super-Hubble regime.

The additional exponential factor in the power spectrum of the tensor modes is a very important difference. It will appear crucial later when we will discuss the tensor to scalar ratio $r$.

Also here $k$ denotes the length of the comoving spatial momentum, which in turn originates from the spatial Fourier transform. The latter has the form of the plane wave $e^{i\vec k\vec x}$. We follow the standard convention that indices of $k_i$ are raised and lowered by the $3$-dimensional spatial metric that is $\delta_{ij}$ in our consideration.

\subsection{Scalar perturbations}
We start by noting that for sure using the spatially Fourier transformed quantities is very handy. We therefore proceed working with all fields being spatially Fourier transformed. Such fields are function of $\tau$ and the comoving spatial momentum $\vec k$, such as $\varphi(\tau,\vec k)$. We recite here 
the following crucial notation from (\ref{frwboxsk}):
\begin{equation}
	\bar \Box_k\varphi(\tau,\vec k)=-\frac1{a^2}(\pd_0^2+2\Hc\pd_0+k^2)\varphi(\tau,\vec k)  \, .
	\label{frwboxskmaintext}\end{equation}
Note that we use $\pd_0\equiv\pd/\pd\tau$.

\subsubsection{Linear variation of EOM for scalar perturbations}

We here use standard notations. In particular the 
perturbed line element reads
\begin{equation}
ds^2=a(\tau)^2\left[-(1+2\phi)d\tau^2-2\partial_i \beta d\tau 
dx^i+((1-2\psi)\delta_{ij}+2\partial_i\partial_j\gamma)dx^idx^j\right]\, . 
\label{mFr}
\end{equation}
We immediately move to gauge-invariant variables putting aside all possible gauge-fixing issues. 
For the scalar perturbations these variables are known as Bardeen potentials 
\cite{Bardeen:1980kt,Mukhanov:1990me} defined as follows, 
\begin{equation}
\Phi=\phi-\frac{1}{a}(a\vartheta)^\prime=\phi-\dot 
\chi\, ,\quad\Psi=\psi+\Hc\vartheta=\psi+H\chi \, ,
\label{GIvars}
\end{equation}
where  $\chi=a\beta+a^2\dot\gamma$, $\vartheta=\beta+\gamma'$, $\Hc(\tau)=a'/a$, and prime hereafter denotes the derivative with respect to the conformal time $\tau$. One more useful gauge-invariant quantity is defined by
\begin{eqnarray}
\delta R_{\rm GI}&=& \delta R - \bar R'(\beta+\gamma') =\nonumber \\
&=& 2(\bar R+3\bar\Box_k)\Psi-2\bar R(\Phi+\Psi)-6\frac{a'}{a^3}(\Phi'+\Psi')+2\frac{k^2}{a^2}(\Phi+\Psi) \, .
\end{eqnarray}
The variation of the trace equation (\ref{tEOMtrace}) reads 
\begin{eqnarray}
\delta E=-2\lambda\Pc\zeta =0 \, ,\label{eq3410}
\end{eqnarray}
where
\begin{eqnarray}
 \zeta &=&  \delta\Box \bar R+(\bar\Box_k-r_1)\delta R_{\rm GI} \, ,\nonumber\\
 \Pc &\equiv& \LT \partial^{\mu}\bar R\partial_{\mu} - \frac2{a^2}\LF \bar R'' + 2\Hc \bar R'\RF\RT 
\frac{\Fc\LF\bar \Box_k\RF - \Fc_1}{\LF\bar \Box - r_1\RF^2} + 
3\Fc\LF\bar \Box_k\RF%\nonumber\\
+ (\bar R + 3r_1)\frac{\Fc\LF\bar \Box_k\RF - \Fc_1}{\bar \Box_k - r_1}\, ,
\nonumber\\
 \delta\Box &=&  \frac{1}{a^2}\left[2\Phi\left(\partial_0^2+2\Hc
\partial_0\right)+\left(\Phi'+3\Psi'\right)\partial_0\right] \, .\nonumber
\end{eqnarray}
We notice that $\zeta$ actually is the variation of the ansatz relation (\ref{ansatz}), and in the local $R^2$ gravity theory $\zeta=0$. The fractions with denominators containing operator $\bar \Box_k$ are still analytic functions. This can be shown by using the Taylor series expansion of $\Fc(\bar \Box_k)$ in the numerators near the point $r_1$ and the required condition $\Fc^{(1)}(r_1)=0$. Equation (\ref{eq341}) is exactly the same as in \cite{Biswas:2012bp} because the additional term containing the Weyl tensor enters the trace equation only as a $O({\bf C}^2)$ term and thus does not appear in perturbations around a conformally flat background.

Even though the expression for $\delta E$ is manifestly homogeneous with respect to $\zeta$ it may be convenient to rewrite it in the following different way, 
\begin{equation}
	\!\!\!
	\delta E=\frac{2\lambda}{a^2}\left[\bar R'\Xi' +2\LF \bar R'' + 2\Hc \bar R'\RF\Xi-a^2(\bar R+3r_1)(\Upsilon-\Fc_1\delta R_{\rm GI})-3a^2\Fc(\bar \Box_k)\zeta\right] = 0 \, ,
	\label{eq341}
\end{equation}
where
\begin{eqnarray} 
\Upsilon=\frac{\Fc(\bar \Box_k)-\Fc_1}{\bar\Box_k-r_1}\zeta+\Fc_1\delta R_{\rm GI} 
\: \text{ and } \: 
\Xi=\frac{\Fc(\bar\Box_k)-\Fc_1}{(\bar \Box_k-r_1)^2}\zeta \, .\nonumber
\end{eqnarray}
The variation of the $({}^i_j)$-equation with $i\neq j$ in the system (\ref{tEOM}) yields
\begin{equation}
	\delta E^i_j=-2\lambda\frac{k^ik_j}{a^2}\left[\Fc_1(\bar R+3r_1)(\Phi-\Psi)+\Upsilon\right]+2\lambda c^i_j=0 \, . 
\label{eq342}
\end{equation}
Here we have introduced the following notation, 
\begin{equation}
	c^\mu_\nu=(\bar R^\alpha_\beta+2{\bar\nabla_k}^{\alpha}{\bar\nabla_k}{}_{\beta})\Fc_{\rm C}(\bar\Box_{Ck})\delta^{(s)} C_{\nu\alpha}^{\phantom{\nu\alpha}\beta\mu} \, . 
	\label{dceom}
\end{equation}
The whole expression above is explained and evaluated explicitly in the appendix~\ref{ap:cmunu}. Moreover, its explicit form follows below after all necessary equations are given. 
Equations (\ref{eq341}) and (\ref{eq342}) are in principle sufficient 
to study the classical dynamics of the perturbations 
because they provide a coupled system of two equations for the two Bardeen potentials. However, one must understand that all equations are in the game, and
we may expect that certain constraints arise
because we have more than two equations for two functions. These constraints must be accounted. This actually happens in the pure GR. 
Furthermore, at least in GR those constraints provide a considerable simplification and they help to write the second variation of the action that we will derive later.

The variation of the $({}^0_i)$-equation in the system (\ref{tEOM}) yields
\begin{equation}
	\!\!\delta E^0_i=2\lambda\frac{ik_i}{a^2}\left[2\Fc_1(\bar R+3r_1)(\Psi'+\Hc\Phi)-(\Upsilon'-\Hc\Upsilon)+\Fc_1\bar R'\Phi-\frac12\bar R'\Xi\right]+2\lambda c^0_i=0 \, . 
	\label{eq343}
\end{equation}

Finally the variation of the $({}^0_0)$-equation in (\ref{tEOM}) 
(perhaps, the most tedious derivation in this section) yields
\begin{eqnarray}
\delta E^0_0&=&  \frac{2\lambda}{a^2}\bigg[-2\Fc_1(\bar R+3r_1)(3\Hc\Psi'+3\Hc^2\Phi+k^2\Psi)+3\Hc\Upsilon'-3\Hc'\Upsilon+k^2\Upsilon \nonumber \\
	& -& 3\Fc_1\bar R'(\Psi'+2\Hc\Phi)-\frac{\bar R'}2\Xi'+\frac12\left({\bar R''}+2\Hc \bar R'\right)\Xi\bigg]+2\lambda c^0_0=0 \, . 
	\label{eq344}
\end{eqnarray}
For completeness we rewrite here from the appendix~\ref{ap:cmunu}, equation (\ref{cmunuap}), the following components of 
$c^\mu_\nu$, 
\begin{eqnarray}
	c^i_j&=& \frac{k^ik_j}{a^2}\left(\Theta'+2\Hc\Theta+(\Hc'-\Hc^2)\Omega+\frac{k^2}3\Omega-a^2\dot H\Omega\right) \: \text{ where } \: i\neq j \, , \nonumber\\
	c^0_i &=&  -\frac23\frac{ik_ik^2}{a^2}\Theta \, , \quad	c^0_0= \frac23\frac{k^4}{a^2}\Omega \, , \nonumber\\
	\Theta&=& \Omega'+2\Hc\Omega\, ,\quad\Omega=\Fc_{\rm C}(\bar \Box_k+6H^2)\frac{\Phi+\Psi}{a^2}\, .\nonumber
	%\label{ccompsmaintext}
\end{eqnarray}
We emphasize that in the definition of $\Omega$ the d'Alembertian operator is the one given in (\ref{frwboxskmaintext}).\footnote{Notice that the present equations (\ref{eq343}), (\ref{eq344}) differ from analogous equations presented in \cite{Biswas:2012bp} in what they must coincide (i.e. without $c^\mu_\nu$) by terms which were irrelevant for the analysis in \cite{Biswas:2012bp}. We assume that it was a misprint type omission in that paper.}

We performed a verification of the equations (\ref{eq341}), (\ref{eq342}), (\ref{eq343}), (\ref{eq344}) by evaluating the Bianchi identities in appendix~\ref{ap:bianchi}. One more equation, namely the $({}^i_i)$ component with no sum over $i$ is needed for this verification and it is presented in the appendix~\ref{ap:bianchi} as well. 
Such an equation is not necessary for the main computations here although may be useful elsewhere.

\subsubsection{de Sitter limit of linearized equations for scalar perturbations}
%%%
The dS limit is what actually must be analyzed to get the inflationary observables. Moreover, assuming a nearly dS background we greatly simplify the perturbation equations.
Equation (\ref{eq342}) multiplied by $k^2$ reduces to
\begin{eqnarray}
	-k^2\Fc_1(\bar R+3r_1)(\Phi-\Psi)-k^2\Upsilon+k^2\left(\Theta'+2\Hc\Theta+\frac{k^2}3\Omega\right)=0 \, ,
\label{eq342ds}
\end{eqnarray}
equation (\ref{eq343}) multiplied by $3\Hc$ reduces to
\begin{equation}
	6\Hc\Fc_1(\bar R+3r_1)(\Psi'+\Hc\Phi)-3\Hc(\Upsilon'-\Hc\Upsilon)-2\Hc k^2\Theta=0 \, , 
	\label{eq343ds}
\end{equation}
while equation (\ref{eq344}) reduces to
\begin{eqnarray}
	-2\Fc_1(\bar R+3r_1)(3\Hc\Psi'+3\Hc^2\Phi+k^2\Psi)+3\Hc\Upsilon'-3\Hc'\Upsilon+k^2\Upsilon+\frac23k^4\Omega=0 \, .
	\label{eq344ds}
\end{eqnarray}
Summing up all three equations above, using that in dS limit $\Hc'=\Hc^2$, expanding $\Theta$ in terms of $\Omega$, using the explicit representation for $\Omega$ in terms of Bardeen potentials and cancelling common $k^2$ factor we end up with 
\begin{equation}
	[\Fc_1(\bar R+3r_1)+(\bar \Box_k-2H^2)\Fc_{\rm C}(\bar \Box_k+6H^2)]\frac{\Phi+\Psi}{a^2}=0 \, .
	\label{phipsiweyl}
\end{equation}
From here it is obvious that, whether there is no Weyl tensor contribution in the original action (that means $\Fc_{\rm C}(\Box)=0$), the following very neat condition arises: $\Phi+\Psi=0$. 

At this moment we observe the very crucial fact. The non-local operator in the latter equation (\ref{phipsiweyl}) is nothing but the operator $\Pc(\bar\Box)$ defined in (\ref{newroots}) provided we identify the operators $\bar\Box_k+2H^2$ and $\bar \Box$, take the approximation $r_1\ll\bar R$, and factor out the constant multiplier $\Fc_1r_1$. This observation immediately guarantees that in a general situation even with the Weyl tensor term in the model the only solution to equation (\ref{phipsiweyl}) reads 
\begin{equation}
	\Phi+\Psi=0 \, . 
	\label{mainzero}
\end{equation}
This in turn has the following dramatic consequences:
\begin{equation}
	c^\mu_\nu=\Omega=\Theta=0 \, , 
	\label{mainzeroc}
\end{equation}
meaning that the Weyl term in the action does not influence the scalar perturbations during the nearly dS expansion at all.

Furthermore, in the dS limit we have
\begin{eqnarray}
\zeta=(\bar \Box_k-r_1)\delta R_{\rm GI} \,  , \quad \Upsilon=\Fc(\bar \Box_k)\delta R_{\rm GI}\, ,
\end{eqnarray}
and the perturbation of the trace equation (\ref{eq341}) simplifies to
\begin{equation}
	\Fc(\bar \Box_k)(\bar R+3\bar \Box_k)\delta R_{\rm GI}=(\bar R+3\bar \Box_k)\Upsilon=\Fc_1(\bar R+3r_1)\delta R_{\rm GI} \, . 
	\label{dtraceds}
\end{equation}
Explicitly replacing $\delta R_{\rm GI}$ in the RHS of the latter equation yields
\begin{equation}
	\Upsilon=2\Fc_1(\bar R+3r_1)\Psi \, . 
	\label{dtracedsmore}
\end{equation}
Notice that our results for the constraints coincide with those obtained in \cite{Craps:2014wga} (as it should be as the presence of Weyl-dependent term in the action was just proven not to influence the scalar perturbations in the nearly dS phase).

\subsubsection{Quadratic variation of action for scalar perturbations}

Computing the complete quadratic variation around an arbitrary solution in our non-local model seems  a very hard task. The main and only obstacle is the $R\Fc(\Box)R$ term. The variation of the Weyl tensor term is generic for any conformally flat background.
Even though we consider already special backgrounds satisfying a simplifying ansatz (\ref{ansatz}) (which made it possible to compute the complete linearized equations of motion) we leave the problem of computing  the complete second order action variation for the future work. For the present paper we concentrate on questions 
only requiring to compute the second order variation around a nearly dS background. 

We start with the variation of the Einstein-Hilbert action that we already prepared in (\ref{deltaEHfull}) and 
we consider the scalar perturbations using the following line element, 
\begin{equation}
ds^2=a(\tau)^2\left[-(1+2\Phi)d\tau^2+(1-2\Psi)\delta_{ij}dx^idx^j\right] \, , 
\label{mFraction}
\end{equation}
The substitution of this simple line element in (\ref{delta0lambda}) is considerably long, but he result upon omitting total derivative terms,  reads
\begin{equation}
\begin{split}
	\delta_{\rm EH}^{(s)}=&\frac{1}{a^2}\left[-6{\Psi'}^2+6k^2\Psi^2-12\Hc\Psi'(\Phi+\Psi)-9\Hc^2(\Phi+\Psi)^2-4k^2\Psi(\Phi+\Psi)\right]\, , \\
	\delta_g^{(s)}=&-\frac12\left[\Phi^2+6\Phi\Psi-3\Psi^2\right]  \, . 
\end{split}
\label{d2GRscalar}
\end{equation}
Here the superscript $(s)$ stands for scalars. This result is valid for any scale factor and one can check that it coincides with the result presented in \cite{Mukhanov:1990me}.

Next we make the variation of the operator quadratic in the scalar curvature. Doing this around dS spacetime we write
\begin{equation}
	\delta^2 S_{R^2}^{(s)}=\int d\tau d\vec k\sqrt{|\bar g|}\frac{\lambda}{2}\left[(2\delta_{\rm EH}^{(s)}-\bar R \delta_g^{(s)})f_0\bar R+\delta R_{\rm GI}\Fc(\bar\Box_k)\delta R_{\rm GI}\right] \, , 
	\label{d2RFRscalards}
\end{equation}
which differs from (\ref{d2RFRfullds}) in using $\delta^{(s)}$ instead of $\delta$ and by the use of $\delta R_{\rm GI}$ in place of a generic $\delta R$. 

The last piece we need is the variation of the operator quadratic in the Weyl tensor, which can be immediately written using the results in appendix~\ref{ap:cmunu}, namely 
\begin{equation}
	\delta^2 S_{C^2}^{(s)}=\int d\tau d\vec k\sqrt{|\bar g|}\frac{\lambda}{2}\frac{4k^4}3\frac{\Phi+\Psi}{a^2}\Fc_{\rm C}(\bar\Box_k+6H^2)\frac{\Phi+\Psi}{a^2} \, , 
	\label{d2C2scalar}
\end{equation}
where the factor $4k^4/3$ comes from the full contraction of two $K$-tensors also defined in appendix~\ref{ap:cmunu}. This particular result is valid for a general scale factor. Furthermore, thanks to the constraint (\ref{mainzero}) this contribution vanishes around the nearly dS inflationary phase.

Now technically we have all pieces of the second variation for the non-local action around a nearly dS background.

\subsubsection{Quantization of scalar perturbations}

We must derive an effective action for a canonical variable that in turn has to be cooked up in order to make use of the quantization procedure described in \cite{Mukhanov:1990me}. Generically it is a tedious task,  but in this paper we consider only the nearly dS phase of inflation and, therefore, we quantize the perturbations in a nearly dS background. Moreover, we have proven that the Weyl tensor contribution for the scalar perturbations goes away and we just come to the configuration considered in \cite{Craps:2014wga}.
Therefore, we do not have to do anything rather than list the results obtained previously. Those results absolutely support the statement that the non-local modification of $R^2$ theory retains the same value for the scalar power spectrum as a local $R^2$ theory as long as the nearly dS phase of evolution is considered.

In particular the following quantity was analyzed, 
\begin{equation}
	\Rc=\Psi+\frac{H}{\dot{\bar R}}\delta R_{\rm GI}  \, , 
	\label{isoR}
\end{equation}
which during the nearly dS phase can be approximated by
\begin{equation}
	\Rc\approx\frac{H^2}{\dot H}\Psi \, . 
	\label{isoRds}
\end{equation}
Moreover, this quantity enjoys the conservation in the slow-roll inflation in the super-Hubble limit. 
Then the power spectrum during the crossing of the Hubble radius is: 
\begin{equation}
	|\delta_\Rc(\tau,\vec k)|^2\approx\frac{H^6_{k=Ha}}{16\pi^2\dot H^2_{k=Ha}}\frac{1}{3\lambda\Fc_1\bar R} \, , 
	\label{isoRps}
\end{equation}
which is manifestly a scale invariant expression.

\subsection{Tensor to scalar ratio $r$}

This ratio is simply defined as the ratio of power spectra of tensor and scalar perturbations. 
At the moment of the Hubble radius crossing (when (\ref{pshexpsub}) and (\ref{pshexpsuper}) coincide) we get
\begin{equation}
	r=\frac{2|\delta_h|^2}{|\delta_\Rc|^2}=48\frac{\dot H^2}{H^4}e^{2\omega(\bar R/6)}\, ,
	\label{rexp}
\end{equation}
where the factor 2 accounts for two polarizations of the tensor modes. Next using the connection between the slow-roll parameter $\epsilon_1=-(\dot H/H^2)|_{k=Ha}$ and the number of $e$-foldings $N$, namely
\begin{equation*}
	N=\int_{t_i}^{t_f}Hdt=\frac1{2\epsilon_1} \,  ,
\end{equation*}
we arrive at
\begin{equation}
	r=48\epsilon_1^2e^{2\omega(\bar R/6)}=\frac{12}{N^2}e^{2\omega(\bar R/6)}\, ,
	\label{rexpepsN}
\end{equation}
which is the familiar result (see \cite{DeFelice:2010aj}) modulo the exponential factor.

\section{Conclusions}

In this paper we have presented a 
class of weakly non-local gravitational theories in the Weyl basis. These theories provide a super-renormalizable or finite UV completion of the Einstein-Hilbert theory as well as the local $R^2$ theory.
The theory presented here is very general and all the freedom is encoded in the choice of two
form-factors (entire functions). Here we selected two of them compatible with stability (no tachyons) and unitarity (no ghosts).
However, we did not investigate in this paper the whole landscape of 
theories likely compatible with general principles. This sounds like a very important question to determine possible classes of form-factors for which the resulting theory is healthy, i.e. unitary and UV-complete.

Next, we explicitly showed that any solution of a local $R^2$ gravity is a solution in our non-local framework. This is including a situation without a cosmological constant. To embed a solution of a local $R^2$ gravity in our non-local model one has to fulfill relations (\ref{relnl}). In this case a given solution of a local $R^2$ gravity is an exact solution in our non-local theory. Perturbations, however, must and do differ in local and non-local theories. It is an outstanding mathematical problem to construct explicitly more general analytic solutions beyond this embedding. While some examples are presented in \cite{Dimitrijevic:2015eza}, in general this question is far from being answered in full.

The idea of adopting solutions of a local $R^2$ gravity applies in particular to the $R^2$ inflationary scenario for which all parameter relations can be easily satisfied. This is a very important solution. Indeed, for the moment the accuracy of observational data related to inflation has been increased considerably. It is therefore a test bed for our model. Most of the interesting observables are related to perturbations around the nearly dS expansion during inflation.

We therefore have thoroughly elaborated on linearized equations of motion, quadratic variation of action and quantization of perturbations. It turns out that only the term $R\Fc(\Box)R$ is not yet tamed in full. We can derive its quadratic variation only around a nearly dS background. We hope to extend this to any solution satisfying ansatz (\ref{ansatz}) in our future works. Nevertheless, we managed to deduce all the linearized equations of motion around any background satisfying (\ref{ansatz}), and not only in a dS limit. As the result we found the full quadratic variation of the action around a nearly dS background and quantized scalar and tensor perturbations. A very interesting and useful formula (\ref{hmunugeneric}) for the second variation of a quadratic Weyl tensor term in the action was obtained in passing.

Upon analysis of quantized perturbations we came to a modified ratio of power spectra of tensors and scalars $r$ given in (\ref{rexpepsN}). While other essential parameters, spectral tilts for example, retain their values like in a local $R^2$ gravity the modified $r$ deserves more discussion.  Namely
\begin{equation}
	r_{R^2}=\frac{12}{N^2} \: \text{ vs. } \quad r_{\rm non-local}=\frac{12}{N^2}e^{2\omega\left(\frac{\bar R}{6\Mc}\right)} \, ,
	\label{oldrnewr}
\end{equation}
where we have restored the scale of our theory $\Mc$.
The change is always a positive factor originating from the structure of the non-local operator $\Fc_{\rm C}(\Box)$, yet this factor can be smaller or grater than 1. Explicit expression for $\omega(z)$ comes from (\ref{newroots},\,\ref{newrootsno}). $\bar R$ is the scalar curvature during the nearly dS inflationary phase. 
The function $\omega(z)$ cannot be arbitrarily normalized, but is rather determined from many conditions imposed from various considerations like absence of ghosts in Minkowski background, renormalizability, etc. In other words even though we have freedom to adjust the form-factors, once given, we must accept their respective behaviour. This means that apparently the latter formula can be understood as a one more constraint. More and more accurate measurements of $r$ will give information on how to fix two things: the scale $\Mc$ and the first Taylor coefficients of the form-factors $\gamma_i(z)$ around $z=\bar R/(6\Mc)$. It is currently a work in progress to obtain constraints on $\Mc$ for model form-factors $\gamma_i(z)$ based on the cosmological data.

Existing data are not enough to say somewhat definitive about this new extra multiplier though. Indeed, it is important to have $N\sim 55$ $e$-foldings. This gives $r_{R^2}\approx0.004$ while the most strict upper bound from PLANCK 2015, BICEP2 and Keck Array data \cite{Ade:2015xua,Ade:2015lrj,Array:2015xqh} is $r<0.07$. It is however very important that given that there come incredible improvements in accuracy of the measurements our modified formula $r_{\rm non-local}$ is ready to explain values of $r$ different from a pure local $R^2$ inflation. We also notice that a modification of $r$ established in this paper mathematically resembles the effect of $r$ modification in the so called $\alpha$-attractor inflationary models \cite{Kallosh:2013yoa}. Nevertheless, physically the effects are different. In our present consideration we observe a modified $r$ due to the inclusion of tensorial structures which generate a modified tensor power spectrum. The scalar power spectrum in our setup is preserved at the level of a pure $R^2$ inflation. This automatically preserves the value and scale dependence of $n_s$, the spectral tilt for the scalar power spectrum. For a narrow subclass of inflaton potentials considered in \cite{Kallosh:2013yoa}, namely for the potential $V(\phi)\propto \tanh^2(\alpha\phi)$, it is possible to keep the same scale dependence $n_s(k)$, too. However, this potential is modified compared to the one in the Einstein frame representation of the $R^2$ model.    

It is worth to mention also that our results regarding changes to the tensor perturbations during inflation, in particular a modified value of $r$, seem to have no local counterpart. Given a local curvature squared gravity with $R^2$ and $C_{\mu\nu\alpha\beta}^2$ terms we readily observe 2 poles in the spin-2 propagator (this is read from (\ref{d2spin2dsinf}) with $\Fc_{\rm C}(\Box)=\const$). The second pole will be either ghost or tachyon creating instabilities. Even if someone is ready to live with this, corresponding changes to parameters like $r$ will go way beyond just a simple multiplier. A model of this kind was analyzed in \cite{Khoury:2006fg}. The same should apply to higher but finite number of derivatives as a finite number of poles will be generated. In the non-local gravity however, infinite number of derivatives can be combined to an operator with no poles at all.

There are many open questions of the cosmological origin around obtained here results. One may want to analyze perturbations beyond dS phase. This requires an anticipated quadratic variation of the non-local gravity action around any background (at least satisfying ansatz (\ref{ansatz})). A much more involved problem is a generalization of our considerations to the full inflationary model considered in \cite{Starobinsky:1980te,Starobinsky:1981vz} which uses the local $R^2$ model to get inflation and the graceful exit from it to a matter-dominated stage driven by scalarons at rest, but needs inclusion of one-loop quantum gravitational corrections to obtain decay of scalarons into pairs of particles and anti-particles of all kinds of matter (including the known ones) with their subsequent thermalization leading finally to the hot radiation-dominated stage (the standard hot Big Bang).  Another interesting question is a re-computation of the multipole curve in the non-local theory. While the curve itself finds strong observational support, we can use it to constrain further non-local form-factors. 

As a more ambitious and a very intriguing question we see the development of a model, which could unify bounce and inflation in the non-local gravity framework. We recall that from the point of view of the non-local gravity theories presented in this paper the question of building a model is translated into finding an appropriate form-factors, which allow a given solution to EOM to exist. Enriched, compared to a local $f(R)$ gravity, parameter space in principal allows a co-existence of bounce and inflation. Early studies of this question in a local $R^2$ gravity, where such bounce is possible in closed FLRW backgrounds with a positive spatial curvature, were already done in \cite{Starobinsky:1980te}. Bounce in non-local gravities was studied in, e.g. \cite{Biswas:2005qr,Koshelev:2012qn,Koshelev:2013lfm,Dimitrijevic:2015eza,Dimitrijevic:2015eaa} and many interesting and important results are already uncovered. Nevertheless, a complete scenario joining bounce and inflation in one is still missing.

Returning to the present paper we conclude that for the moment we have provided a class of gravity theories, which accommodate at the same time unitarity, stability, asymptotic polynomial behaviour of the form-factors, and inflation. We can successfully confront inflationary parameters being derived in our model with their respective observed values.

\acknowledgments
AK is supported by the FCT Portugal fellowship SFRH/BPD/105212/2014 and in part by FCT Portugal grant UID/MAT/00212/2013 and by RFBR grant 14-01-00707. 
AS was supported by the RSF grant 16-12-10401.

\appendix

\section{Notations}
\label{ap:notation}
%\subsection{Common notations}

We use $\kappa_D$ for the gravitational coupling and $M_P$ for the Planck mass. $D$ is the dimension. 
Unless specified explicitly we work in $D=4$. In this case we have
\begin{equation}
	\frac1{2\kappa_4^{2}}=\frac{M_P^2}{2}=\frac{1}{16\pi G_N}\, ,
\end{equation}
where $G_N$ is the Newtonian constant. The metric signature is:
\begin{eqnarray} g_{\mu\nu}=(-,+,+,+,\dots),\quad g_{\mu\nu}g^{\mu\nu}=D\, .
\end{eqnarray}
The $4$-dimensional indices are labelled by small Greek letters. The metric-compatible 
connection (Christoffel symbols) reads 
\begin{eqnarray}
\Gamma_{\mu\nu}^\rho=\frac12g^{\rho\sigma}(\pd_\mu g_{\nu\sigma}+\pd_\nu
g_{\mu\sigma}-\pd_\sigma g_{\mu\nu}) \, . 
\end{eqnarray}
The covariant derivative is denoted by $\nabla_\mu$ and acts as follows, 
\begin{eqnarray}
\cpd_\mu F^{.\alpha.}_{.
\beta.}=\pd_\mu F^{.\alpha.}_{.
\beta.}+\Gamma^\alpha_{\mu\chi}F^{.\chi.}_{.
\beta.}-\Gamma^\chi_{\mu\beta}F^{.\alpha.}_{.\chi.} \, .
\end{eqnarray}
It follows that $\nabla_\rho g_{\mu\nu}\equiv0$.
The Riemann tensor, curvatures and the Einstein tensor are defined as
\begin{eqnarray}
 R^\sigma_{\mu\nu\rho}&=& \pd_\nu\Gamma^\sigma_{\mu\rho}
-\pd_\rho\Gamma^\sigma_{\mu\nu}+\Gamma^\sigma_{\chi\nu}\Gamma^\chi_{\mu\rho}
-\Gamma^\sigma_{\chi\rho}\Gamma^\chi_{\mu\nu}\,
,\quad R_{\mu\rho}=R^\sigma_{\mu\sigma\rho}\, ,\quad R=R^\mu_\mu \, , \\
G_{\mu\nu}&=& R_{\mu\nu}-\frac12Rg_{\mu\nu} \, . 
\end{eqnarray}
The commutator of covariant derivatives and the d'Alembertian (box) operators are
\begin{eqnarray}
[\cpd_\mu,\cpd_\nu]A_\rho=R^\chi_{\rho\nu\mu}A_\chi \, , \quad \Box=g^{\mu\nu}\cpd_\mu\cpd_\nu \, . 
\end{eqnarray}
The Weyl tensor follows from  the Ricci decomposition, namely 
\begin{eqnarray}
 C^\mu_{\alpha\nu\beta}&=&  R^\mu_{\alpha\nu\beta} -\frac 1{D-2}(\delta^\mu_\nu
R_{\alpha\beta}-\delta^{\mu}_\beta
R_{\alpha\nu}+{g}_{\alpha\beta}R^\mu_\nu-{g}_{\alpha\nu}R^\mu_\beta
)\nonumber\\
&+&\frac R{(D-2)(D-1)} (\delta^\mu_\nu {g}_{\alpha\beta}-\delta^{\mu}_\beta
{g}_{\alpha\nu}) \, . 
 \end{eqnarray}
The Weyl tensor has all the symmetry properties of the
Riemann tensor and it is absolutely traceless, i.e.
\begin{eqnarray} 
C^\mu_{\alpha\mu\beta}=0 \, . 
\end{eqnarray}
Moreover it is invariant under Weyl rescalings, i.e.
\begin{eqnarray}
\hat C^\mu_{\alpha\beta\gamma} = C^\mu_{\alpha\beta\gamma} \: \text{ for } \: \hat g_{\mu\nu} = \Omega^2(x) g_{\mu\nu} \, . 
\end{eqnarray}
The latter implies that the Weyl tensor is zero for conformally flat manifolds 
(i.e. those manifolds, where the metric can be brought to the form 
$ds^2=a(x)^2\eta_{\mu\nu}dx^\mu dx^\nu$ with
$\eta_{\mu\nu}$ being the Minkowski metric with the same signature).

When the index structure is not crucial we use
$${\bf R}\:\text{ for }\: R\, ,\quad{\bf Ric}\:\text{ for }\: R_{\mu\nu}\, ,\quad{\bf Riem}\:\text{ for }\: R_{\mu\nu\alpha\beta}\, ,\quad{\bf C}\:\text{ for }\: C_{\mu\nu\alpha\beta}\, ,\quad$$
for brevity.

The bar is used to designate background quantity upon perturbations such that
$$\varphi=\bar\varphi+\delta\varphi\, .$$

%\subsection{FLRW Spacetimes}
The FLRW Universe is described by the following metric, 
\begin{eqnarray}
	ds^2=-dt^2+a(t)^2\left(\frac{dr^2}{1-Kr^2}+r^2d\Omega^2\right) \, .
	\label{frwcosmic}
\end{eqnarray}
Here $t$ is the cosmic time and $a(t)$ is the scale factor. 
$K=\pm1,0$ designates open, closed and spatially flat configurations. 
In the present paper we focus on the spatially flat case with $K=0$. 
Then we can write the FLRW metric as
\begin{equation}
	ds^2=-dt^2+a(t)^2\delta_{ij}dx^idx^j \, .
	\label{frwcosmicflat}
\end{equation}
The Hubble function is defined as $H=\dot a/a$ with dot denoting the derivative with respect to $t$.

Equivalently one can rewrite (\ref{frwcosmicflat}) as
\begin{eqnarray}
	ds^2=a(\tau)^2\left(-d\tau^2+\delta_{ij}dx^idx^j\right)\, .\label{frwtau}
\end{eqnarray}
Here $\tau$ is the conformal time and the relation between it and the cosmic time is $ad\tau=dt$. Hence, the FLRW Universe is conformally flat and the Weyl tensor in it is identically zero. The background quantities in the latter metric are
\begin{eqnarray}
	\Gamma^\mu_{0\nu}&=& \Hc\delta^\mu_\nu \, , \quad \Gamma^0_{\mu\nu}=\delta_{\mu\nu}\Hc\, ,\quad\Hc=a'/a \, ,\\
	R&=& \frac{6}{a^2}(\Hc'+\Hc^2) \, , \quad R_{\mu\nu} 
=\left( 
	\begin{array}{cc}
		-3\Hc'&0\\
		0&(\Hc'+2\Hc^2)\delta_{ij}
	\end{array}
	\right)\, ,\\
	R^0_{i0j}&=& \Hc'\delta_{ij} \, , \quad R^i_{0j0}=-\Hc'\delta^i_{j} \, , \quad R^i_{jkm} =\Hc^2(\delta^i_k\delta_{jm}-\delta^i_m\delta_{kj}) \, . 
	\label{frwtaucscurves}
\end{eqnarray}
We use the index ``$0$'' for the tau component of any tensor (the cosmic time is used less often and wherever needed we designate it with the index $t$). Latin small letters from the middle of the alphabet  
are used for the spatial indices, while with $'$ we denote the derivative with respect to the conformal time $\tau$. 
We finally introduce the following useful relations, 
\begin{eqnarray}
 	\pd_0&=& a\pd_t \, , \quad H=\Hc/a \, , \nonumber\\
	\Box_S&=& -\frac{1}{a^2}(\pd_0^2+2\Hc\pd_0-\delta^{ij}\pd_i\pd_j) \, , 
	\label{frwtaubox} \\
	R&=& 6\dot H+12 H^2 \, , \nonumber
\end{eqnarray}
where $\Box_S$ is the d'Alembertian operator acting on scalars.

Perturbation functions are not space-homogeneous. It is convenient to perform a spatial Fourier transform which for some function $\varphi(\tau,\vec x)$ in the case $K=0$ reads
$$
\varphi(\tau,\vec x)=\int\varphi(\tau,\vec k)e^{i\vec k\vec x}d\vec k \, .
$$
Using the above Fourier representation one readily observes that
\begin{equation*}
	\bar \Box_S\varphi(\tau,\vec x)=-\int\frac1{a^2}(\pd_0^2+2\Hc\pd_0+k^2)\varphi(\tau,\vec k)d\vec k 
	\, , 
\end{equation*}
where the bar means that the scalar d'Alembertian operator is not perturbed.
The following notation is extremely useful:
\begin{equation}
	\bar \Box_k\varphi(\tau,\vec k)=-\frac1{a^2}(\pd_0^2+2\Hc\pd_0+k^2)\varphi(\tau,\vec k)  \, .
	\label{frwboxsk}\end{equation}

%\subsection{de Sitter Spacetime}
In the dS limit we have:
\begin{eqnarray}
	H&\approx& \const  \, , \quad 
	a\approx a_0e^{Ht} \, , \\
	R^\sigma_{\mu\nu\rho}&\approx& \frac {R}{12}(\delta^\sigma_\nu g_{\mu\rho}-\delta^\sigma_\rho g_{\mu\nu}) \, , \quad  
	R^\mu_\nu
	\approx\frac{R}4\delta^\mu_\nu,\quad
	R\approx 12H^2\approx\const.
	\label{nearlyds}
\end{eqnarray}
In the conformal time we observe that
\begin{equation}
	a\approx-\frac{1}{H\tau}\, ,\quad\Hc\approx-\frac{1}{\tau}\:\text{ and }\:\Hc'\approx\Hc^2\approx\frac{1}{\tau^2}\, .
	\label{nearlydstau}
\end{equation}

\section{Derivation of equations (\ref{starinffactor}), (\ref{starH}) and (\ref{starR})}\label{apstarobinsky}

For the Lagrangian density (\ref{R2action}), the $(00)$-component of the gravitational field equations for an isotropic 
homogeneous spatially flat universe, a spatially flat FLRW background, is
the following third order differential equation with respect to the scale factor $a(t)$:
\begin{eqnarray}
\frac{H\dot R}{M^2}&-&3(\dot H +H^2)\left(1+\frac{R}{3M^2}\right)+\frac{1}{2}\left(R+\frac{R^2}{6M^2}\right)=
{\kappa_4^2\rho} \,  ,
\label{R2eqn}
\end{eqnarray}
Further, we consider the stage of expansion $H>0$. In the absence of any kind of matter ($\rho=0$), it has two symmetries: 
invariance under time translation and scale factor dilatation. Thus, it can be reduced to the first order equation \cite{Starobinsky:1980te}:
\begin{equation}
\frac{dy}{dx} = -\frac{M^2}{12x^{1/3}y} - 1 \, , \quad x=H^{3/2}\, ,\quad y=\frac{\dot H}{2H^{1/2}}\,  ,\quad
dt=\frac{dx}{3x^{2/3}y}\, .
\label{1stordereq}
\end{equation}

The inflationary regime $|\dot H|\ll H^2$ corresponds to $y<0,~|y|\ll x,~|dy/dx|\ll 1$. Thus, in the leading order of the slow-roll 
approximation:
\begin{equation}
y=y_0=-\frac{M^2}{12x^{1/3}}\,  , \quad x\gg M^{3/2}\, ,\quad t_s-t = \frac{6x^{2/3}}{M^2}=\frac{6H}{M^2}\, .
\end{equation}
Let us calculate the next order correction: $y=y_0+y_1$. Then
\begin{equation}
\frac{M^2 y_1}{12x^{1/3}y_0^2}=\frac{dy_0}{dx}=\frac{M^2}{36x^{4/3}} \, .
\end{equation}
From here,
\begin{equation}
y_1=\frac{y_0^2}{3x}=\frac{M^4}{432x^{5/3}}\, .
\end{equation}
The next order correction to $y$ is proportional to $x^{-3}$. It follows from this that
\begin{equation}
\dot H=2yx^{1/3}= -\frac{M^2}{6}+\frac{M^4}{216x^{4/3}}=-\frac{M^2}{6}+\frac{M^4}{216H^2}=-\frac{M^2}{6}+
\frac{1}{6(t_s-t)^2}\, ,
\label{dotH}
\end{equation}
where the zero-order result for $H$ is used to obtain the last term. Integration of (\ref{dotH}) leads to
 (\ref{starinffactor}), (\ref{starH}), (\ref{starR}) with $r_1=M^2$.

\section{Evaluation of $c^\mu_\nu$ in equation (\ref{dceom})}
\label{ap:cmunu}
Just a brute force computation gives
\begin{equation}
	\delta^{(s)} C_{\nu\alpha}^{\phantom{\nu\alpha}\beta\mu}=\frac{\Phi+\Psi}{a^2} K_{\nu\alpha}^{\phantom{\nu\alpha}\beta\mu} \, , 
	\label{deltacnl}
\end{equation}
where $\delta^{(s)}$ means that the variation is evaluated on scalar perturbations. $K_{\nu\alpha}^{\phantom{\nu\alpha}\beta\mu}$ is a constant tensor whose components depend on $k_i$ only. Moreover it is absolutely traceless and retains all the symmetry properties of the Weyl tensor. Its explicit form is
\begin{eqnarray}
 K_{0j}^{\phantom{0j}0i}&=& -\frac16k^2\delta^i_j+\frac12k^ik_j \, , \nonumber\\
K_{mj}^{\phantom{mj}ki}&=& \frac13k^2(\delta^k_m\delta^i_j-\delta^k_j\delta^i_m)
-\frac12\delta^k_mk^ik_j
-\frac12\delta^i_jk^kk_m
+\frac12\delta^k_jk^ik_m
+\frac12\delta^i_mk^kk_j  \,  .\nonumber
\end{eqnarray}
Few comments are in order. First, variation of the Weyl tensor is manifestly gauge-invariant in our case, because this tensor is zero on the background. Second, again because the Weyl tensor is zero on the background its variation must retain the property to be trace-free.\footnote{Our formula for the Weyl tensor variation differs from the one presented in \cite{Durrer:2004fx}. Namely \cite{Durrer:2004fx} claims that only $(0i0j)$ components are non-trivial (for which our results coincide). We stress that the formula for the variation of the Weyl tensor presented in \cite{Durrer:2004fx} does not provide a traceless tensor as it should.} Third, the appearance of a factor $\Phi+\Psi$ is easy to understand as the case $\Phi+\Psi=0$ corresponds to a specific form of the metric perturbation such that $h_{\mu\nu}=2\Phi g_{\mu\nu}$. One easily sees here a conformal rescaling of the metric. The Weyl tensor in turn is known to transform covariantly (rescaling by the conformal function) under conformal transformations. It is even conformally invariant provided that one of its indices is up and others are down. The latter implies that for $\Phi+\Psi=0$ combined with the fact that the Weyl tensor is zero on the background the variation of the Weyl tensor must be zero as it is obvious from equation (\ref{deltacnl}).

The next step is to compute
\begin{equation}
	\bar\Box_{Ck}\delta^{(s)} C_{\nu\alpha}^{\phantom{\nu\alpha}\beta\mu}=\left[(\bar \Box_k+6H^2)\frac{\Phi+\Psi}{a^2}\right] K_{\nu\alpha}^{\phantom{\nu\alpha}\beta\mu} \, .
	\label{boxdeltacnl}
\end{equation}
We pay attention to the fact that $\bar\Box_{Ck}$ is a new operator. Bar as before indicates that it is not perturbed, the subscript $C$ reminds that it is the d'Alembertian operator with all accompanying connection terms due to the rank of the tensor on the right, the subscript $k$ means that the tensor on the right is Fourier transformed with respect to the spatial coordinates. The essential result here is that on the right hand side box acts on a scalar function only. This results through a recursion relation in
\begin{equation}
	\Fc_{\rm C}(\bar\Box_{Ck})\delta^{(s)} C_{\nu\alpha}^{\phantom{\nu\alpha}\beta\mu}=\left[\Fc_{\rm C}(\bar \Box_k+6H^2)\frac{\Phi+\Psi}{a^2}\right] K_{\nu\alpha}^{\phantom{\nu\alpha}\beta\mu} \, .
	\label{fcboxdeltacnl}
\end{equation}
To finalize the evaluation of (\ref{dceom}) we have to compute
\begin{equation}
	c^\mu_\nu=(\bar R^\alpha_\beta+2{\bar\nabla_k}^\alpha{{\bar\nabla}_k}{}_\beta)\Omega(\tau)K_{\nu\alpha}^{\phantom{\nu\alpha}\beta\mu} \: 
	\text{ where } \: \Omega(\tau)=\Fc_{\rm C}(\bar \Box_k+6H^2)\frac{\Phi+\Psi}{a^2} \, . 
\end{equation}
The subscript $k$ in $\nabla$ technically means that wherever encountered, we replace $\pd_i\to ik_i$. Some long, careful and rather tough computation gives
\begin{equation}
	\begin{split}
	c^i_j=&\frac{k^ik_j}{a^2}\left(\Omega''+4\Hc\Omega'+(3\Hc'+3\Hc^2)\Omega+\frac{k^2}3\Omega-a^2\dot H\Omega\right)
	\text{ where } \: i\neq j   \,  ,  \\
	c^0_i=&-\frac23\frac{ik_ik^2}{a^2}(\Omega'+2\Hc\Omega)\, ,\quad
	c^0_0 = \frac23\frac{k^4}{a^2}\Omega \, ,\\ 
	c^i_i=&-\frac{1}{3a^2}\left((k^2-3k^ik_i)(\Omega''+4\Hc\Omega'+(3\Hc'+3\Hc^2+a^2\dot H)\Omega)+{k^2}(k^2-k^ik_i)\Omega\right) \, ,
\end{split}
\label{cmunuap}
\end{equation}
with no sum over $i$ in the last expression. First we stress that the above result is for a generic scale factor $a(\tau)$. As a verification of the above formulae one can check that the trace $c^\mu_\mu$ vanishes. Also we notice that the terms proportional to $\dot H$ originate from the Ricci tensor term. Hence, this contributions cancel in the dS limit when $H$ is almost a constant. This is very much correct, since the Ricci tensor becomes proportional to the metric tensor and being contracted with the $K$-tensor results in the zero trace.

\section{Verification of equations (\ref{eq341}), (\ref{eq342}), (\ref{eq343}), (\ref{eq344})}
\label{ap:bianchi}
We verify the equations for perturbations by evaluating the Bianchi identity. The identity dictates that
\begin{eqnarray}
\nabla_\mu G^\mu_\nu\equiv0 \, .
\end{eqnarray}
This implies that in any generally covariant gravity modification EOM should form a rank-2 tensor obeying Bianchi identity. Therefore, for system (\ref{tEOM}) we can write
\begin{eqnarray}
\nabla_\mu E^\mu_\nu=0 \, .
\end{eqnarray}
This fact at the background level was explicitly verified in \cite{Koshelev:2013lfm,Biswas:2013cha}. At the perturbed level up to the linear order we must have
\begin{equation}(\delta\nabla_\mu) {\bar E}^\mu_\nu+{\bar \nabla}_\mu(\delta E^\mu_\nu)=0\, .\end{equation}
However, to simplify the task, we will assume that we do the computation around a given background, meaning that ${\bar E}^\mu_\nu=0$ even though the identical cancellation should hold in general. This simplification is valid since we were already using the ansatz (\ref{ansatz}) and relation among parameters (\ref{relnl}) in derivation of the perturbation equations. Having said this we must show that
\begin{eqnarray}
\bar \nabla_\mu(\delta E^\mu_\nu)=0 \, , 
\end{eqnarray}
for all $\nu$. Moreover, since this is an identity it must hold for any values of constant parameters in $\delta E^\mu_\nu$. This in turn implies that it must hold independently for the $c^\mu_\nu$ tensor. We thus start by showing that
\begin{equation}
\bar\nabla_\mu c^\mu_\nu=0\label{bidc} \, . 
\end{equation}
Accounting (\ref{cmunuap}) and performing a lengthy algebra one can demonstrate explicitly that (\ref{bidc}) holds. This in the meantime verifies (\ref{cmunuap}) itself.

The next step is to verify that
\begin{equation}
{\bar\nabla}_\mu(\delta E^\mu_\nu(\Fc_{\rm C}=0))=0 \, . 
\label{bienodc}
\end{equation}
For this aim we need an explicit form of $\delta E^i_i$, where no sum is taken and $\Fc_{\rm C}=0$. This is given as
\begin{eqnarray}
	&\delta E^i_i&(\Fc_{\rm C}=0)=   \nonumber\\
	&=& -\frac{2\lambda}{a^2}\Fc_1(\bar R+3r_1)\left[ (\Psi-\Phi)(k^2-k^ik_i)+2(\Psi''+2\Hc\Psi')+2(\Hc\Phi'+(\Hc^2+2\Hc')\Phi) \right] \nonumber\\
	&-&\frac{2\lambda}{a^2}\Hc\Upsilon'+\left( \frac{2\lambda}{a^2}(2\Hc'+\Hc^2-k^ik_i)-\lambda (\bar R+2r_1) \right)\Upsilon+\frac\lambda{a^2}(\bar R'\Xi'+(\bar R''+2\Hc\bar R')\Xi)  
	\nonumber\\
	&+&\frac{2\lambda}{a^2}\Fc_1\bar R'(\Psi'+2\Hc\Phi)-2\lambda\Fc(\bar\Box_k)\zeta \, .
	\label{eqii}
\end{eqnarray}
As a first check for the latter equation itself one can prove that performing the trace $\delta E^\mu_\mu$ one indeed gets (\ref{eq341}). Then after indeed a tedious and long computation one proves that (\ref{bienodc}) is true.

%%%%%%%%%%%%%%%%%%%%%%%%%%%%%%%%%%%%%%%%%%%%%%%%%%%%%%%%%%%%%%%%%%%%%%%%%%%%%%%%%%%%%%%%%
%%%%%%%%%%%%%%%%%%%%%%%%%%%%%%%%%%%%%%%%%%%%%%%%%%%%%%%%%%%%%%%%%%%%%%%%%%%%%%%%%%%%%%%%%

%\bibliography{s}{}
%\bibliographystyle{JHEP}
\providecommand{\href}[2]{#2}\begingroup\raggedright\endgroup

\end{document}